\documentclass[journal]{IEEEtran}
\usepackage{array}
\usepackage{amsmath, amssymb, bm}
\usepackage{graphicx}
\usepackage{cite}
\usepackage{algorithm}
\usepackage{algorithmic}
\usepackage{multirow}
\usepackage{capt-of}
\usepackage{tabularx}
\usepackage{url}
\usepackage{hyperref}
\usepackage{xcolor} 
\usepackage[normalem]{ulem}
\usepackage[section]{placeins}
\usepackage{tikz}
\usetikzlibrary{positioning,arrows.meta,shapes.geometric,fit,calc}
\usepackage{tikz-cd}
\usepackage{multicol}
\begin{document}

\title{Physics-Aware Transformer-Based Beamforming Optimization for Near-Field E-MIMO ISAC Systems}

\author{
Mohsin Ali,~\IEEEmembership{Member,~IEEE}, Abdulahi Abiodun Badrudeen,~\IEEEmembership{Member,~IEEE}, Azharul Islam,~\IEEEmembership{Member,~IEEE}, and Sunwoo Kim,~\IEEEmembership{Senior~Member,~IEEE}
\thanks{Mohsin Ali, Abdulahi Abiodun Badrudeen, Azharul Islam, and Sunwoo Kim are with the Department of Electronic Engineering,
Hanyang University, Seoul, 04763, Republic of Korea (e-mail:
mohsin0987@hanyang.ac.kr; aabadrudeen@hanyang.ac.kr; azharul001@hanyang.ac.kr; remero@hanyang.ac.kr).}}

\markboth{}
{Shell \MakeLowercase{\textit{et al.}}: Bare Demo of IEEEtran.cls for IEEE Journals}

\maketitle
\begin{abstract}
This paper proposes a physics-aware Transformer-based beamforming optimization (PATBO) framework for near-field extreme massive multiple-input multiple-output integrated sensing and communication (E-MIMO-ISAC) systems. PATBO combines regularized zero-forcing (RZF) communication directions and spherical-wave near-field sensing directions with learned stream-power allocation. A token-preserving Transformer maintains the correspondence between individual users/targets and their allocated powers, while a Cram\'er--Rao bound (CRB)-sensitivity-aware objective emphasizes power-allocation errors with greater localization impact. Physical feasibility is explicitly verified after inference, with infeasible predictions handled through physics-constrained power repair and optimizer fallback. For a $32\times32$ E-MIMO array with $1024$ antennas, the proposed structured representation reduces the learned output from $12{,}288$ real-valued coefficients of an unstructured complex beamformer to six stream-power variables. Direct PATBO achieves $100.98\%$ reference-rate retention and $19.54\%$ joint feasibility, which increases to $87.79\%$ after repair. The complete hybrid framework achieves $99.89\%$ reference-rate retention and $100\%$ constraint satisfaction on the retained test set, while requiring optimizer fallback for only $12.21\%$ of the scenarios. Direct PATBO inference requires $1.11$~ms per sample, corresponding to approximately $858\times$ and $779\times$ lower runtime than the evaluated cold-start and PATBO-warm-start sequential quadratic programming (SQP) implementations, respectively.
\end{abstract}


\section{Introduction}
\label{sec:introduction}

Integrated sensing and communication (ISAC) is a key enabling
technology for future sixth-generation (6G) wireless networks,
allowing communication and environmental sensing to share spectrum,
hardware, and waveform resources \cite{Liu2022ISAC,Ali2026Joint}. In parallel,
extreme massive multiple-input multiple-output (E-MIMO) provides large
beamforming gains, high spatial resolution, and enhanced spatial
multiplexing capability \cite{Wang2024XLMIMO, Lu2024NearFieldTutorial}. Its enlarged physical aperture, however, significantly extends the radiative near-field region, where spherical-wave propagation introduces range-dependent spatial characteristics and enables joint angular--range beam focusing \cite{Ouyang2024Primer,Li2024Modular}. These properties make near-field E-MIMO particularly attractive for simultaneous
communication, sensing, and localization.

Recent studies have investigated near-field ISAC beamforming,
localization, Cram\'er--Rao bound (CRB) analysis, wideband operation, and complexity-aware designs \cite{Wang2024CRB}, \cite{Qu2024NearFieldISAC}, \cite{Ray2025ISAC}, \cite{Li2024ISPAC}, \cite{Zhao2024NearFieldISAC}, \cite{Wang2024WidebandNFISAC}, \cite{Lin2025NFISACComplexity}. Related developments include modular extremely large-scale multiple-input multiple-output
(XL-MIMO) hybrid beamforming, movable-antenna-assisted near-field ISAC,
and practical ultra-wideband architectures
\cite{Meng2025HybridNFISAC}, \cite{Ding2026MovableNFISAC}, \cite{Kim2026NearFieldChallenges}. Near-field beam acquisition, channel
estimation, beam training, and localization have also received
considerable attention \cite{Lei2025NFLocalizationSurvey}, \cite{Zeinali2026Beamforming},\cite{Jiang2023DLBeamTraining}, \cite{Shi2025DoubleSided}, \cite{Yang2025NFLocalization}. These works demonstrate the benefits of near-field propagation, while also highlighting the increased complexity of beam management in large-aperture systems.

Near-field E-MIMO-ISAC requires the joint coordination of communication and sensing resources. Communication users require
sufficient signal-to-interference-plus-noise ratio (SINR) and achievable rate, whereas sensing targets require adequate sensing SINR and localization accuracy. Existing ISAC beamforming methods therefore commonly employ iterative numerical optimization \cite{Cui2025EnergyISAC},\cite{Zhao2025Multifunctional}, \cite{Zhang2024ActiveSTARISAC}, \cite{hua2025near}, \cite{RobustNFISAC2026}, \cite{Mao2024CellFreeRegion}, \cite{Demirhan2025CellFreeISAC}. Although such methods provide effective benchmarks, repeatedly solving constrained nonconvex problems for changing user and target geometries can incur substantial online latency, particularly for E-MIMO systems with large-dimensional beamforming matrices.

Learning-based beamforming offers an alternative by mapping channel
or geometry information directly to beamforming variables
\cite{Xia2020DLBeamforming}, \cite{Zhang2020DLBeamforming}. Structured
learning methods have further addressed constrained beamforming using penalty-dual, Transformer, graph-based, and related architectures \cite{li2024learning}, \cite{Li2024HPETransformer}. Learning-assisted ISAC beamforming has also been investigated through predictive, deep-unfolding, Transformer, and graph-learning approaches \cite{DeepUnfoldingRISISAC}, \cite{Gao2026Deep},\cite{PredictiveComplex2024}, \cite{Yu2025PredictiveISAC},
\cite{Zhang2026AI}, \cite{Jiang2026GraphISAC}. These studies demonstrate the potential of learning to reduce the repeated computational burden of iterative beamforming optimization. However, most existing learning-based designs either predict high-dimensional beamforming variables directly or focus primarily on communication and sensing performance without explicitly addressing the strong localization sensitivity introduced by near-field CRB constraints. The joint consideration of near-field E-MIMO geometry, low-dimensional physically structured communication--sensing beamforming, CRB-sensitive learning,
and explicit post-inference feasibility recovery therefore remains
insufficiently addressed.

A further challenge is that prediction accuracy and physical
feasibility are not equivalent in constrained ISAC beamforming.
Accurately approximating a reference optimization solution does not
necessarily guarantee satisfaction of coupled nonconvex constraints
\cite{li2024learning}, \cite{Li2024HPETransformer}. This issue becomes
particularly important in near-field localization, where the CRB
depends on the transmit covariance, geometry-dependent steering-vector derivatives, and the resulting Fisher information matrix. Consequently, relatively small stream-power prediction errors can produce noticeable CRB degradation when the operating point lies close to the localization-feasibility boundary. A learning framework should therefore account not only for prediction accuracy but also for the physical consequences of prediction errors.

Direct prediction of the complete complex-valued beamforming matrix
also becomes increasingly inefficient as the E-MIMO array size grows. For an $M$-element array serving $K$ communication users and $L$ sensing targets, direct beamformer prediction requires
$M(K+L)$ complex, or equivalently $2M(K+L)$ real-valued, output
variables. In contrast, exploiting physically meaningful communication and sensing directions allows the learning task to be
restricted to the $K+L$ stream-power variables. Such a structured
representation reduces the learning dimension while retaining the
near-field spatial characteristics through physics-based beam
construction.

To address these challenges, this paper proposes a
physics-aware Transformer-based beamforming optimization
(PATBO) framework for near-field E-MIMO-ISAC systems. PATBO constructs regularized zero-forcing (RZF) communication directions and near-field sensing directions from the system geometry, while the Transformer predicts only the $K+L$ normalized stream-power fractions. Geometry-aware entity tokens are processed using a token-preserving Transformer architecture that models inter-entity interactions without discarding the association between each user or target and its corresponding power variable. Safe-margin reference generation is employed to construct robust training targets, while CRB-sensitivity weighting emphasizes power-allocation errors having a stronger impact on localization
performance. After inference, physical feasibility is explicitly
verified; infeasible predictions are passed to a low-dimensional
power-repair stage, and optimizer fallback is invoked when repair is unsuccessful.

The main contributions of this work are summarized as follows:
\begin{itemize}

\item We develop a structured near-field E-MIMO-ISAC beamforming
framework that combines RZF communication directions and
spherical-wave near-field sensing directions with learned stream-power
allocation. For $M=1024$, $K=4$, and $L=2$, this reduces the learned
output representation from $12{,}288$ real-valued beamforming
coefficients to only six stream-power variables.

\item We propose a token-preserving Transformer with
CRB-sensitivity-aware training. The architecture preserves the
association between individual users/targets and their allocated
powers, while emphasizing prediction errors having greater impact on
near-field localization performance.

\item We develop a hierarchical verify--repair--fallback strategy for
physical constraint satisfaction. Physics-constrained power repair
increases joint feasibility from $19.54\%$ to $87.79\%$, while the
complete hybrid framework achieves $100\%$ feasibility on the retained
test set with optimizer fallback required for only $12.21\%$ of the
samples.

\end{itemize}

The remainder of this paper is organized as follows.
Section~\ref{sec:system_model} presents the near-field E-MIMO-ISAC
system model and joint beamforming problem. Section~\ref{sec:patbo} introduces the proposed PATBO framework. Section~\ref{sec:results} the numerical results, and Section~\ref{sec:conclusion} concludes the paper.

\textit{Notation:}
Boldface lowercase and uppercase letters denote vectors and matrices, respectively. The superscripts $(\cdot)^T$ and $(\cdot)^H$ denote the transpose and Hermitian transpose, respectively. The operators $\|\cdot\|_2$ and $\|\cdot\|_F$ denote the Euclidean and Frobenius norms, while $|\cdot|$ denotes the absolute value of a scalar. $\mathbf{I}_N$ denotes the $N\times N$ identity matrix.
The sets of real and complex matrices of size $M\times N$ are denoted by $\mathbb{R}^{M\times N}$ and $\mathbb{C}^{M\times N}$,
respectively. The operators $\operatorname{tr}(\cdot)$ and
$\operatorname{diag}(\cdot)$ denote the matrix trace and diagonal
matrix construction, respectively, while $\Re\{\cdot\}$ denotes the
real part of a complex-valued quantity.

\section{System Model and Problem Formulation}
\label{sec:system_model}

\subsection{Near-Field E-MIMO-ISAC System}

We consider a downlink E-MIMO-ISAC system in which a base station (BS) equipped with an $M$-element antenna array simultaneously serves $K$ single-antenna communication users and senses $L$ targets. The communication and sensing functionalities share the same transmit array, bandwidth, and transmit-power budget.

Let
\begin{equation}
    \mathbf{W}
    =
    \left[
    \mathbf{w}_{c,1},\ldots,\mathbf{w}_{c,K},
    \mathbf{w}_{s,1},\ldots,\mathbf{w}_{s,L}
    \right]
    \in
    \mathbb{C}^{M\times(K+L)}
    \label{eq:joint_beamforming_matrix}
\end{equation}
denote the joint transmit beamforming matrix, where $\mathbf{w}_{c,k}$ is the beamforming vector associated with communication user $k$, and $\mathbf{w}_{s,l}$ denotes the sensing beam associated with target $l$.

The transmitted signal is
\begin{equation}
    \mathbf{x}
    =
    \sum_{k=1}^{K}
    \mathbf{w}_{c,k}s_{c,k}
    +
    \sum_{l=1}^{L}
    \mathbf{w}_{s,l}s_{s,l},
    \label{eq:transmitted_signal}
\end{equation}
where $s_{c,k}$ and $s_{s,l}$ denote the normalized communication and sensing symbols, respectively. Assuming mutually uncorrelated
unit-power streams, the total transmit power is $P_{\mathrm{tx}}=\|\mathbf{W}\|_F^2$. Accordingly, the transmit-power constraint is
\begin{equation}
    \|\mathbf{W}\|_{F}^{2}
    \leq
    P_t,
    \label{eq:power_constraint}
\end{equation}
where $P_t$ denotes the maximum available transmit power.

\subsection{Near-Field Propagation Model}

Due to the large physical aperture of the E-MIMO array, users and
sensing targets may lie inside the radiative near-field region.
Therefore, spherical-wave propagation is considered instead of the
conventional plane-wave approximation \cite{Cui2022ChannelEstimation}.

Let $\mathbf{a}_m\in\mathbb{R}^{3}$ denote the Cartesian position of antenna element $m$, and let $\mathbf{p}=[x,y,z]^T$ denote the
position of a user or sensing target. The propagation distance between antenna $m$ and position $\mathbf{p}$ is $d_m(\mathbf{p})=\|\mathbf{p}-\mathbf{a}_m\|_2$.
A spherical-wave near-field response can be expressed as
\begin{equation}
    [\mathbf{a}(\mathbf{p})]_m
    =
    \frac{1}{\sqrt{4\pi d_m(\mathbf{p})}}
    \exp
    \left(
    -j\frac{2\pi}{\lambda}d_m(\mathbf{p})
    \right),
    \label{eq:nearfield_response}
\end{equation}
where $\lambda$ denotes the carrier wavelength. The amplitude term
accounts for distance-dependent propagation attenuation, whereas the spherical-wave phase preserves the element-dependent propagation distance required for near-field focusing. Consequently, both range and angular information are retained in the array response \cite{Wu2024HierarchicalBeam}.

For communication user $k$, let $\mathbf{h}_k\in\mathbb{C}^{M\times1}$ denote the corresponding channel vector. The overall communication channel matrix is
\begin{equation}
    \mathbf{H}
    =
    [\mathbf{h}_1,\ldots,\mathbf{h}_K]
    \in
    \mathbb{C}^{M\times K}.
    \label{eq:channel_matrix}
\end{equation}

\subsection{Communication Performance}

The received signal of communication user $k$ can be written as
\begin{equation}
\begin{split}
    y_k
    ={}&
    \mathbf{h}_k^{H}\mathbf{w}_{c,k}s_{c,k}
    +
    \sum_{\substack{i=1\\i\neq k}}^{K}
    \mathbf{h}_k^{H}\mathbf{w}_{c,i}s_{c,i}
    \\
    &+
    \sum_{l=1}^{L}
    \mathbf{h}_k^{H}\mathbf{w}_{s,l}s_{s,l}
    +
    n_k,
\end{split}
\label{eq:received_signal}
\end{equation}
where $n_k$ denotes additive noise with variance $\sigma_n^2$.
Accordingly, the communication SINR is
\begin{equation}
    \gamma_k^{c}
    =
    \frac{
    |\mathbf{h}_k^{H}\mathbf{w}_{c,k}|^2
    }{
    \displaystyle
    \sum_{\substack{i=1\\i\neq k}}^{K}
    |\mathbf{h}_k^{H}\mathbf{w}_{c,i}|^2
    +
    \sum_{l=1}^{L}
    |\mathbf{h}_k^{H}\mathbf{w}_{s,l}|^2
    +
    \sigma_n^2
    }.
    \label{eq:communication_sinr}
\end{equation}
The achievable rate of user $k$ is
\begin{equation}
    R_k
    =
    \log_2(1+\gamma_k^{c})
    \quad \text{bit/s/Hz},
    \label{eq:user_rate}
\end{equation}
and the system sum rate is
\begin{equation}
    R_{\mathrm{sum}}
    =
    \sum_{k=1}^{K}R_k.
    \label{eq:sum_rate}
\end{equation}
Each communication user is required to satisfy
$R_k\geq R_{\min}$, $k=1,\ldots,K$.

\subsection{Near-Field Sensing and Localization Model}
\label{subsec:sensing_localization}

In addition to serving the $K$ communication users, the E-MIMO
transmitter simultaneously illuminates $L$ sensing targets. Let
$\mathbf{p}_l=[x_l,y_l,z_l]^T$, $l=1,\ldots,L$, denote the
three-dimensional Cartesian position of sensing target $l$. Owing to the electrically large aperture of the E-MIMO array, the sensing
targets may lie in the radiative near-field region. Therefore, the
sensing model is constructed using position-dependent near-field
steering vectors rather than conventional far-field angle-only
steering vectors \cite{Wang2024CRB,Hua2024NearField3D}.

For target $l$, let $\mathbf{a}_l\triangleq\mathbf{a}(\mathbf{p}_l)
\in\mathbb{C}^{M\times1}$ denote its near-field steering vector.
The corresponding normalized near-field focusing direction is
constructed as
\begin{equation}
    \mathbf{v}_{s,l}
    =
    \frac{
        \mathbf{a}(\mathbf{p}_l)
    }{
        \|\mathbf{a}(\mathbf{p}_l)\|_2
    }.
    \label{eq:normalized_nf_sensing_beam}
\end{equation}
Unlike far-field beam steering, which primarily controls the angular direction of the transmitted beam, \eqref{eq:normalized_nf_sensing_beam} exploits the spherical-wave
propagation structure to focus the transmitted energy toward a
specific three-dimensional position \cite{Li2024Modular,Wu2024HierarchicalBeam}.

\subsubsection{Sensing SINR}

The sensing quality of target $l$ is characterized by its sensing
SINR. The dedicated sensing beam $\mathbf{w}_{s,l}$ constitutes the desired sensing signal, whereas the sensing beams intended for the other targets are treated as inter-target sensing interference. Accordingly, the sensing SINR of target $l$ is given by
\begin{equation}
    \gamma_l^{s}
    =
    \frac{
        \left|
        \mathbf{a}_l^{H}
        \mathbf{w}_{s,l}
        \right|^2
    }{
        \displaystyle
        \sum_{\substack{j=1\\j\neq l}}^{L}
        \left|
        \mathbf{a}_l^{H}
        \mathbf{w}_{s,j}
        \right|^2
        +
        \sigma_n^2
    },
    \label{eq:sensing_sinr}
\end{equation}
where $\sigma_n^2$ denotes the sensing-noise power. In the considered sensing model, interference among the dedicated sensing streams is explicitly accounted for in \eqref{eq:sensing_sinr}.

To guarantee a minimum sensing quality for every target, the
target-wise sensing-SINR constraint is
\begin{equation}
    \gamma_l^{s}
    \geq
    \gamma_{\min}^{s},
    \qquad
    l=1,\ldots,L,
    \label{eq:sensing_sinr_constraint}
\end{equation}
where $\gamma_{\min}^{s}$ denotes the minimum required sensing-SINR
threshold. For performance evaluation, the worst-case sensing SINR
over all targets is defined as
\begin{equation}
    \gamma_{\mathrm{wc}}^{s}
    =
    \min_{l=1,\ldots,L}
    \gamma_l^{s}.
    \label{eq:worst_case_sensing_sinr}
\end{equation}
This metric determines whether all sensing targets simultaneously
satisfy the required sensing-SINR condition.

\subsubsection{Near-Field Localization CRB}

In addition to the sensing-SINR requirement, localization accuracy
is quantified through the CRB. Near-field
sensing with extremely large arrays provides range-dependent
geometric information that can be exploited for three-dimensional
target localization, whose theoretical estimation accuracy can be
characterized through the Fisher information matrix (FIM) and CRB
\cite{Hua2024NearField3D}.

The target location is specified by three Cartesian coordinates.
Define the Jacobian of the near-field steering vector with respect to the target position as
\begin{equation}
    \mathbf{G}_l
    =
    \left[
    \frac{\partial\mathbf{a}(\mathbf{p}_l)}{\partial x_l},
    \frac{\partial\mathbf{a}(\mathbf{p}_l)}{\partial y_l},
    \frac{\partial\mathbf{a}(\mathbf{p}_l)}{\partial z_l}
    \right]
    \in\mathbb{C}^{M\times3}.
    \label{eq:steering_jacobian}
\end{equation}

In the numerical implementation, the spatial derivatives in
\eqref{eq:steering_jacobian} are evaluated using the central
finite-difference approximation
\begin{equation}
    \frac{
        \partial\mathbf{a}(\mathbf{p}_l)
    }{
        \partial[\mathbf{p}_l]_q
    }
    \approx
    \frac{
        \mathbf{a}
        \left(
        \mathbf{p}_l+\delta\mathbf{e}_q
        \right)
        -
        \mathbf{a}
        \left(
        \mathbf{p}_l-\delta\mathbf{e}_q
        \right)
    }{
        2\delta
    },
    \label{eq:numerical_jacobian}
\end{equation}
where $\mathbf{e}_q$ denotes the $q$th Cartesian basis vector and
$\delta=10^{-4}$~m is the perturbation step used in the numerical
evaluation.

For a given joint beamforming matrix $\mathbf{W}$, define the
transmit covariance matrix as $\mathbf{Q}=\mathbf{W}\mathbf{W}^{H}$. The FIM associated with the Cartesian position of target $l$ is then evaluated as
\begin{equation}
    \mathbf{J}_l
    =
    \frac{
        2|\alpha_l|^2
    }{
        \sigma_n^2
    }
    \Re
    \left\{
        \mathbf{G}_l^{H}
        \mathbf{Q}
        \mathbf{G}_l
    \right\},
    \label{eq:fim_position}
\end{equation}
where $\alpha_l$ denotes the effective target-reflection coefficient. The FIM is made numerically symmetric before evaluating the localization bound.

The Cartesian localization CRB for target $l$ is defined as
\begin{equation}
    C_l(\mathbf{W})
    =
    \operatorname{tr}
    \left(
        \mathbf{J}_l^{-1}
    \right).
    \label{eq:target_crb}
\end{equation}
Hence, a smaller value of $C_l(\mathbf{W})$ corresponds to greater
position information and, consequently, a lower theoretical
localization-error bound.

For robust numerical evaluation of \eqref{eq:target_crb}, let
$\lambda_{l,i}$, $i=1,2,3$, denote the eigenvalues of the symmetrized FIM, with $\lambda_{l,\max}=\max_{i=1,2,3}\lambda_{l,i}$. A relative eigenvalue floor is introduced as $\lambda_{\mathrm{floor},l} =\epsilon_{\lambda}\lambda_{l,\max}$, where $\epsilon_{\lambda}=10^{-12}$ in the numerical implementation. The effective eigenvalues are then defined as
\begin{equation}
    \widetilde{\lambda}_{l,i}
    =
    \max
    \left\{
        \lambda_{l,i},
        \lambda_{\mathrm{floor},l}
    \right\}.
    \label{eq:safe_eigenvalues}
\end{equation}
Accordingly, the numerically robust CRB used in the simulations is
evaluated as
\begin{equation}
    \widetilde{C}_l(\mathbf{W})
    =
    \sum_{i=1}^{3}
    \frac{1}{
        \widetilde{\lambda}_{l,i}
    }.
    \label{eq:robust_crb}
\end{equation}
For the $L$ sensing targets, the aggregate localization metric is
defined as
\begin{equation}
    C(\mathbf{W})
    =
    \sum_{l=1}^{L}
    \widetilde{C}_l(\mathbf{W}).
    \label{eq:aggregate_crb}
\end{equation}
The aggregate localization accuracy is constrained according to
\begin{equation}
    C(\mathbf{W})
    \leq
    \Gamma_{\mathrm{CRB}},
    \label{eq:crb_constraint}
\end{equation}
where $\Gamma_{\mathrm{CRB}}$ denotes the maximum allowable aggregate CRB.

Consequently, the sensing component of the considered E-MIMO-ISAC
system is governed by two complementary requirements:
\eqref{eq:sensing_sinr_constraint} guarantees sufficient target-wise sensing SINR, whereas \eqref{eq:crb_constraint} controls the three-dimensional localization accuracy. These two sensing requirements are incorporated jointly with the communication and transmit-power constraints in the beamforming optimization problem formulated next.

\subsection{Joint Beamforming Problem}

The general beamforming design objective is to maximize the communication sum rate while satisfying the communication QoS, sensing SINR, localization CRB, and total transmit-power requirements, consistent with the constrained beamforming formulations commonly considered for near-field ISAC systems \cite{Qu2024NearFieldISAC, Lin2025NFISACComplexity}. Before imposing the structured beam-direction parameterization adopted in this work, this design objective can be expressed in terms of the joint beamforming matrix as
\begin{subequations}
\label{eq:original_problem}
\begin{align}
    \underset{\mathbf{W}}{\operatorname{maximize}}
    \quad&
    \sum_{k=1}^{K}R_k
    \label{eq:original_objective}
    \\
    \text{subject to}\quad&
    R_k\geq R_{\min},
    \quad
    \forall k,
    \label{eq:original_qos}
    \\
    &
    \gamma_l^{s}
    \geq
    \gamma_{\min}^{s},
    \quad
    \forall l,
    \label{eq:original_sensing}
    \\
    &
    C(\mathbf{W})
    \leq
    \Gamma_{\mathrm{CRB}},
    \label{eq:original_crb}
    \\
    &
    \|\mathbf{W}\|_F^2
    \leq
    P_t.
    \label{eq:original_power}
\end{align}
\end{subequations}

Directly optimizing \eqref{eq:original_problem} over an unrestricted complex beamforming matrix would involve $2M(K+L)$ real-valued coefficients. In the proposed framework, however, both the numerical reference and PATBO employ the structured beam directions introduced in Section~\ref{sec:patbo}. Consequently, the implemented optimization problem is reduced to the $K+L$ stream-power variables associated with these directions. PATBO further replaces repeated numerical optimization of these powers with learned prediction, followed by explicit physical verification and selective feasibility recovery.

\section{Proposed Physics-Aware Transformer-Based Beamforming Optimization}
\label{sec:patbo}

\begin{figure*}[!t]
\centering
\renewcommand{\small}{\fontsize{12}{13}\selectfont}
\renewcommand{\scriptsize}{\fontsize{12}{13}\selectfont}
\resizebox{\textwidth}{!}{%
\begin{tikzpicture}[
    font=\rmfamily\small,
    every node/.style={align=center},
    >=Latex,
    mainbox/.style={
        draw,
        rounded corners=3pt,
        minimum width=3.25cm,
        minimum height=9cm,
        text width=2.85cm,
        align=center,
        line width=0.8pt
    },
    innerbox/.style={
        draw,
        rounded corners=3pt,
        align=center,
        inner sep=5pt,
        line width=0.5pt
    },
    flow/.style={
        ->,
        line width=0.9pt
    },
    decision/.style={
        diamond,
        draw=blue!70!black,
        line width=0.8pt,
        aspect=1.7,
        align=center,
        inner sep=1pt,
        font=\scriptsize
    }
]

\definecolor{myblue}{RGB}{20,60,190}
\definecolor{mygreen}{RGB}{20,100,55}
\definecolor{mypurple}{RGB}{65,30,180}
\definecolor{myorange}{RGB}{220,65,20}
\definecolor{myred}{RGB}{205,25,20}

\node[
    mainbox,
    draw=myblue
] (scenario) {};

\node[
    anchor=north,
    text=myblue,
    font=\small\bfseries
] at ([yshift=-0.25cm]scenario.north)
{Near-field\\scenario};

\node at ([yshift=0.85cm]scenario.center)
{
    \includegraphics[width=2.5cm]{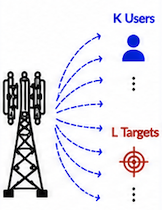}
};

\node[
    innerbox,
    draw=blue!20,
    fill=blue!3,
    text width=2.55cm,
    anchor=south,
    font=\small
] at ([yshift=0.20cm]scenario.south)
{Near-field spherical-wave propagation from BS to $K$ users and $L$ targets.};

\node[
    mainbox,
    draw=mygreen,
    right=0.30cm of scenario
] (token) {};

\node[
    anchor=north,
    text=mygreen,
    font=\small\bfseries
] at ([yshift=-0.25cm]token.north)
{Geometry-aware\\tokenization};

\node at ([yshift=0.85cm]token.center)
{
    \includegraphics[width=2.5cm]{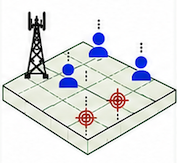}
};

\node[
    innerbox,
    draw=green!20,
    fill=green!3,
    text width=2.55cm,
    anchor=south,
    font=\small
] at ([yshift=0.20cm]token.south)
{Encodes antenna, user, and target geometry using near-field spatial features.};

\node[
    mainbox,
    draw=mypurple,
    right=0.30cm of token
] (patbo) {};

\node[
    anchor=north,
    text=mypurple,
    font=\small\bfseries
] at ([yshift=-0.25cm]patbo.north)
{PATBO\\Transformer};

\coordinate (C) at ([yshift=0.85cm]patbo.center);

\foreach \ang in {30,90,150,210,270,330}{
    \coordinate (n\ang) at ($(C)+(\ang:1.05cm)$);
    \draw[mypurple,line width=0.55pt] (C) -- (n\ang);
    \fill[white] (n\ang) circle (0.10cm);
    \draw[mypurple,line width=0.8pt] (n\ang) circle (0.10cm);
}

\foreach \a in {30,90,150,210,270,330}{
    \foreach \b in {30,90,150,210,270,330}{
        \draw[mypurple!45,line width=0.25pt] (n\a) -- (n\b);
    }
}

\node[
    draw=mypurple,
    rounded corners=2pt,
    fill=white,
    font=\small\bfseries,
    minimum width=1.3cm,
    minimum height=0.55cm
] at (C) {PATBO};

\node[
    innerbox,
    draw=purple!20,
    fill=purple!3,
    text width=2.55cm,
    anchor=south,
    font=\small
] at ([yshift=0.20cm]patbo.south)
{Learns geometry-aware interactions across $K$ users and $L$ targets.};

\node[
    mainbox,
    draw=myorange,
    right=0.30cm of patbo
] (power) {};

\node[
    anchor=north,
    text=myorange,
    font=\small\bfseries
] at ([yshift=-0.25cm]power.north)
{Normalized\\power allocation};

\node at ([yshift=1cm]power.center)
{
    \includegraphics[width=3cm]{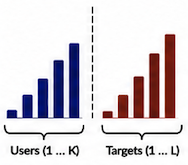}
};

\node[
    innerbox,
    draw=orange!25,
    fill=orange!3,
    text width=2.55cm,
    anchor=south,
    font=\small
] at ([yshift=0.20cm]power.south)
{Transformer outputs normalized power allocation for communication and sensing streams.};

\node[
    mainbox,
    draw=myblue,
    right=0.30cm of power
] (beam) {};

\node[
    anchor=north,
    text=myblue,
    font=\small\bfseries,
] at ([yshift=-0.25cm]beam.north)
{Structured\\beamformer\\(physics-based)};

\node at ([yshift=0.85cm]beam.center)
{
    \includegraphics[width=3cm]{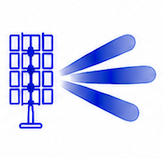}
};

\node[
    innerbox,
    draw=blue!20,
    fill=blue!3,
    text width=2.55cm,
    anchor=south,
    font=\small
] at ([yshift=0.20cm]beam.south)
{RZF precoding for users and near-field focusing for targets construct the beamformer.};

\node[
    mainbox,
    draw=black,
    minimum width=3.6cm,
    right=0.30cm of beam
] (check) {};

\node[
    anchor=north,
    font=\small\bfseries
] at ([yshift=-0.25cm]check.north)
{Feasibility\\check};

\draw[line width=0.9pt,rounded corners=1pt]
([xshift=-0.40cm,yshift=1.35cm]check.center)
rectangle
([xshift=0.40cm,yshift=2.25cm]check.center);

\draw[line width=0.8pt]
([xshift=-0.16cm,yshift=2.35cm]check.center)
rectangle
([xshift=0.16cm,yshift=2.18cm]check.center);

\foreach \yy in {2.00,1.75,1.50}{
    \node[font=\small] at
    ([xshift=-0.17cm,yshift=\yy cm]check.center) {$\checkmark$};
}

\node[innerbox,text width=2.90cm,font=\small] (qos) at (check.center)
{Communication QoS};

\node[innerbox,text width=2.90cm,font=\small,below=0.1cm of qos] (sinr){Sensing SINR};

\node[
    innerbox,
    text width=2.90cm,
    font=\small,
    below=0.1cm of sinr
] (crb) {Aggregate CRB};

\node[
    innerbox,
    text width=2.90cm,
    font=\small,
    below=0.1cm of crb
]  (pwr) {Transmit Power};


\node[
    decision,
    right=0.30cm of check
] (decision1)
{Feasible?};

\node[
    mainbox,
    draw=mygreen,
    right=0.70cm of decision1
] (final) {};

\node[
    anchor=north,
    text=mygreen,
    font=\small\bfseries
] at ([yshift=-0.25cm]final.north)
{Final\\beamformer};

\node at ([yshift=0.85cm]final.center)
{
    \includegraphics[width=3cm]{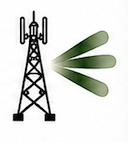}
};
\node[
    innerbox,
    draw=green!20,
    fill=green!3,
    text width=2.55cm,
    anchor=south,
    font=\small
] at ([yshift=0.20cm]final.south)
{Beamformer satisfying all physical constraints.};

\node[
    text=mygreen,
    font=\bfseries,
    above
] at ($(decision1.east)!0.45!(final.west)$)
{Yes};


\draw[flow] (scenario.east) -- (token.west);
\draw[flow] (token.east) -- (patbo.west);
\draw[flow] (patbo.east) -- (power.west);
\draw[flow] (power.east) -- (beam.west);
\draw[flow] (beam.east) -- (check.west);
\draw[flow] (check.east) -- (decision1.west);
\draw[flow](decision1.east) -- (final.west);


\node[
    draw=myred,
    rounded corners=4pt,
    minimum width=2.0cm,
    minimum height=2.05cm,
    text width=2.55cm,
    align=center,
    below=0.7cm of beam,
    line width=0.8pt
] (repair){};

\node[
    anchor=north,
    text=myred,
    font=\small\bfseries
] at ([yshift=-0.25cm]repair.north)
{Physics-\\constrained\\power repair};


\node[
    decision,
    right=1cm of repair
] (decision2)
{Feasible?};

\draw[flow](repair.east) -- (decision2.west);

\draw[flow]
(decision1.south)
-- (decision1.south |- {$(repair.north)+(0,0.35cm)$})
-| (repair.north);

\node[
    text=myred,
    font=\bfseries,
     above
] at ($(decision1.south)+(-0.3cm,-1.5cm)$)
{No};


\node[
    draw=myred,
    rounded corners=4pt,
    minimum width=2.0cm,
    minimum height=2.05cm,
    text width=2.55cm,
    align=center,
    right=1.7cm of decision2,
    line width=0.8pt
] (fallback){};

\node[
    anchor=north,
    text=myred,
    font=\small\bfseries
] at ([yshift=-0.25cm]fallback.north)
{Reference\\optimizer\\fallback};

 \draw[flow](decision2.east)-- (fallback.west);

\node[
    text=myred,
    font=\bfseries,
 above
] at ($(decision2.east)!0.45!(fallback.west)$)
{No};

\draw[flow]
(decision2.south)
-- ++(0,-0.7cm)
-| ([xshift=0cm]final.south);

\node[
    text=mygreen,
    font=\bfseries,
     left
] at ($(decision2.south)+(0cm,-0.4cm)$)
{Yes};

\draw[flow]
(fallback.east)
-- ++(0,0cm)
-| ([xshift=0cm]final.south);

\node[
    draw=mypurple,
    dashed,
    rounded corners=6pt,
    line width=0.8pt,
    minimum width=7.0cm,
    minimum height=2.60cm,
    above=0.45cm of patbo,
    xshift=0cm
] (offline) {};

\node[
    anchor=north,
    text=mypurple,
    font=\small\bfseries
] at ([yshift=-0.15cm]offline.north)
{Offline training};

\node[
    innerbox,
    draw=mypurple,
    minimum width=2.55cm,
    minimum height=0.85cm,
    font=\small
] (margin)
at ([xshift=-1.65cm,yshift=-0.25cm]offline.center)
{Safe-Margin\\Labeling};

\node[
    innerbox,
    draw=mypurple,
    minimum width=2.75cm,
    minimum height=0.85cm,
    font=\small
] (crbloss)
at ([xshift=1.65cm,yshift=-0.25cm]offline.center)
{CRB-Sensitivity\\Aware Loss};

\draw[flow]
(margin.east) -- (crbloss.west);

\draw[flow]
(offline.south) -- (patbo.north);

\end{tikzpicture}%
}
\caption{Proposed PATBO framework for near-field E-MIMO-ISAC combining token-preserving stream-power prediction, physics-based
beamformer reconstruction, feasibility verification, and repair--fallback recovery.}
\label{fig:patbo_framework}
\end{figure*}

\subsection{Overview}

The proposed PATBO framework separates the beamforming problem into
two components: physically constructed beam directions and learned
stream-power allocation. Rather than directly predicting the complete complex-valued beamforming matrix, the Transformer estimates only the normalized powers assigned to the $K$ communication and $L$ sensing streams.

For each scenario, PATBO performs four main operations:
\begin{enumerate}
    \item geometry-aware token construction and Transformer-based
    power prediction;
    \item physical communication and sensing beam-direction generation;
    \item beamformer reconstruction and physical feasibility
    verification; and
    \item adaptive physics-based repair and optimizer fallback when
    required.
\end{enumerate}

The overall architecture of the proposed PATBO framework is
illustrated in Fig.~\ref{fig:patbo_framework}. The near-field
user--target geometry is first represented through geometry-aware
tokens, which are processed by the PATBO Transformer to predict the
normalized power allocation for the communication and sensing
streams. The predicted powers are combined with physically
constructed RZF communication directions and near-field sensing
directions to form the structured beamformer. Its physical
feasibility is then verified with respect to the communication QoS,
sensing SINR, aggregate CRB, and transmit-power constraints.
Infeasible predictions are passed to the physics-constrained
power-repair stage, whereas feasible beamformers are directly
accepted.


\subsection{Geometry-Aware Token Representation}

Each communication user and sensing target is represented by an
individual Transformer token. The communication-user token is
defined as
\begin{equation}
\mathbf{z}_{k}^{c}
=
[
x_k,
y_k,
z_k,
r_k,
\theta_k,
\phi_k,
0,
0,
0,
0
]^T,
\label{eq:comm_token}
\end{equation}
where $(x_k,y_k,z_k)$ denotes the Cartesian position, $r_k$ denotes
the distance from the BS reference point, and $\theta_k$ and $\phi_k$ denote the corresponding angular coordinates.

For sensing target $l$, the token is
\begin{equation}
\mathbf{z}_{l}^{s}
=
[
x_l,
y_l,
z_l,
r_l,
\theta_l,
\phi_l,\\
\sigma_{x,l},
\sigma_{y,l},
\sigma_{z,l},
1
]^T,
\label{eq:sensing_token}
\end{equation}
where $\sigma_{x,l}$, $\sigma_{y,l}$, and $\sigma_{z,l}$ denote the
standard deviations of the assumed target-position uncertainty along the Cartesian axes. The last token entry serves as a type indicator, where zero represents a communication user and one represents a sensing target.
The complete Transformer input is
\begin{equation}
    \mathbf{X}
    =
    [
    \mathbf{z}_{1}^{c},
    \ldots,
    \mathbf{z}_{K}^{c},
    \mathbf{z}_{1}^{s},
    \ldots,
    \mathbf{z}_{L}^{s}
    ]
    \in
    \mathbb{R}^{10\times(K+L)}.
    \label{eq:transformer_input}
\end{equation}

The range is $r=\sqrt{x^2+y^2+z^2}$, and the angles are
$\theta=\operatorname{atan2}(x,z)$ and
$\phi=\allowbreak\operatorname{atan2}\!\left(y,\sqrt{x^2+z^2}\right)$.
For each feature dimension $d$, the normalization parameters $\mu_d$ and $\sigma_d$ are calculated exclusively from the training samples. The normalized feature is
\begin{equation}
    \widetilde{X}_{d,q}
    =
    \frac{X_{d,q}-\mu_d}{\sigma_d}.
    \label{eq:feature_normalization}
\end{equation}
The same training-derived statistics are applied to the validation
and test sets.

\subsection{Token-Preserving Transformer-Based Power Allocation}
\label{subsec:token_preserving_transformer}

To capture the coupled interactions among communication users and sensing targets, PATBO employs a Transformer encoder based on multi-head self-attention. Transformer architectures are well suited to modeling dependencies among sequence elements and have recently been investigated for learning constrained wireless resource-allocation and beamforming mappings \cite{Vaswani2017Attention}.

Each $10$-dimensional entity token is first mapped into a
$d_{\mathrm{model}}=64$ dimensional latent representation. Let
\begin{equation}
\mathbf{Z}^{(0)}
=
\left[
\mathbf{z}^{(0)}_1,
\ldots,
\mathbf{z}^{(0)}_{K+L}
\right]
\in
\mathbb{R}^{d_{\mathrm{model}}\times(K+L)}
\label{eq:embedded_tokens}
\end{equation}
denote the embedded token sequence.

The latent sequence is processed by two Transformer encoder stages, each employing four attention heads and a feed-forward dimension of $128$. A dropout probability of $0.1$ is applied within the network. After the final encoder stage, the resulting token representations are denoted as
\begin{equation}
\mathbf{Z}^{(T)}
=
\left[
\mathbf{z}^{(T)}_1,
\ldots,
\mathbf{z}^{(T)}_{K+L}
\right].
\label{eq:final_token_embeddings}
\end{equation}

Through self-attention, each token incorporates information from all communication users and sensing targets. Hence, the latent representation of a particular stream captures not only its own geometric characteristics, but also the coupled multi-user and multi-target interactions induced by the shared communication, sensing, localization, and transmit-power requirements.

A key feature of the proposed architecture is that the individual token representations are preserved after the Transformer encoders. In particular, no global pooling operation is applied across the token dimension. Instead, each final token representation $\mathbf{z}^{(T)}_q$ is processed using a shared token-wise readout:
\begin{equation}
\ell_q
=
f_{\mathrm{out}}
\left(
\mathbf{z}^{(T)}_q
\right),
\qquad
q=1,\ldots,K+L,
\label{eq:token_logit}
\end{equation}
where $f_{\mathrm{out}}(\cdot)$ consists of a $64$-unit fully connected layer, a Gaussian error linear unit (GELU) activation, dropout, and a scalar output projection.

The resulting $K+L$ entity-specific logits are jointly normalized through a softmax operation:
\begin{equation}
\hat{\rho}_q
=
\frac{
\exp(\ell_q)
}{
\displaystyle
\sum_{j=1}^{K+L}\exp(\ell_j)
},
\qquad
q=1,\ldots,K+L.
\label{eq:token_softmax}
\end{equation}
Therefore, the predicted normalized stream-power vector is
\begin{equation}
\widehat{\boldsymbol{\rho}}
=
[
\hat{\rho}_1,
\ldots,
\hat{\rho}_{K+L}
]^T,
\label{eq:predicted_power_fraction}
\end{equation}
which satisfies
\begin{equation}
\hat{\rho}_q\geq0,
\qquad
\sum_{q=1}^{K+L}\hat{\rho}_q=1.
\label{eq:simplex}
\end{equation}
The corresponding physical stream-power vector is
\begin{equation}
\widehat{\mathbf{p}}
=
P_t\widehat{\boldsymbol{\rho}}.
\label{eq:predicted_physical_power}
\end{equation}

The token-preserving readout is particularly important in the considered problem because each output power corresponds to a specific physical entity. Global aggregation before the output layer can suppress this entity identity, whereas the proposed readout maintains the one-to-one association between the $q$th communication-user or sensing-target token and its corresponding stream-power variable. At the same time, the preceding self-attention layers continue to model interactions among all
entities.

\subsection{Physics-Based Beam-Direction Construction}

The communication beam directions are generated using RZF, which provides a practical tradeoff between multi-user interference suppression and noise enhancement \cite{Peel2005RZF}. The unnormalized RZF direction matrix is
\begin{equation}
    \widetilde{\mathbf{V}}_c
    =
    \mathbf{H}
    \left(
    \mathbf{H}^{H}\mathbf{H}
    +
    \alpha_{\mathrm{RZF}}\mathbf{I}_{K}
    \right)^{-1},
    \label{eq:rzf}
\end{equation}
where the regularization parameter is selected as
\begin{equation}
    \alpha_{\mathrm{RZF}}
    =
    \frac{K\sigma_n^2}{P_{c,0}},
    \qquad
    P_{c,0}=0.8P_t,
    \label{eq:rzf_regularization}
\end{equation}
where $P_{c,0}$ denotes the nominal communication-power level used for constructing the RZF directions in the reference beamformer. Each communication direction is normalized as
\begin{equation}
    \mathbf{v}_{c,k}
    =
    \frac{
    \widetilde{\mathbf{v}}_{c,k}
    }{
    \|\widetilde{\mathbf{v}}_{c,k}\|_2
    }.
    \label{eq:comm_direction}
\end{equation}

For sensing target $l$, the near-field sensing direction is given by the normalized focusing vector in \eqref{eq:normalized_nf_sensing_beam}. Thus, the complete direction
matrix is
\begin{equation}
    \mathbf{V}
    =
    [
    \mathbf{V}_c,\mathbf{V}_s
    ]
    \in
    \mathbb{C}^{M\times(K+L)}.
    \label{eq:direction_matrix}
\end{equation}

\subsection{Beamformer Reconstruction}

The Transformer-predicted beamformer is reconstructed as
\begin{equation}
    \widehat{\mathbf{W}}
    =
    \mathbf{V}
    \operatorname{diag}
    \left(
    \sqrt{
    P_t\widehat{\boldsymbol{\rho}}
    }
    \right).
    \label{eq:beamformer_reconstruction}
\end{equation}
\pagebreak[0]
Since every physical beam direction satisfies $\|\mathbf{v}_q\|_2=1$ and the Transformer output satisfies \eqref{eq:simplex}, it follows that
\begin{equation}
    \|\widehat{\mathbf{W}}\|_F^2
    =
    \sum_{q=1}^{K+L}
    P_t\hat{\rho}_q
    =
    P_t.
    \label{eq:power_guarantee}
\end{equation}
Hence, under the unit-norm beam-direction construction, the
total transmit-power requirement is structurally enforced by
the simplex-constrained output parameterization.

This representation also significantly reduces the learned output
dimension. Direct prediction of the complete complex beamforming matrix requires
\begin{equation}
2M(K+L)
\end{equation}
real-valued outputs, whereas PATBO predicts only $K+L$ real-valued
normalized stream powers. Therefore, the learned output-dimensionality reduction factor relative to an unstructured full-beamformer predictor is
\begin{equation}
\frac{2M(K+L)}{K+L}
=
2M.
\label{eq:dimension_reduction}
\end{equation}

For the considered $M=1024$, $K=4$, and $L=2$ configuration, direct
prediction of the complex beamforming matrix would require $2\times1024\times6 = 12{,}288$ real-valued outputs, whereas PATBO predicts only six normalized stream powers. This corresponds to a $2048\times$ reduction in learned output dimensionality.

This comparison concerns the neural-network output representation.
The numerical reference optimizer employed for label generation also operates over the same $K+L$ structured stream-power variables while keeping the physical communication and sensing beam directions fixed.

\subsection{Safe-Margin Reference Generation}

A neural predictor trained using reference solutions located close to the CRB feasibility boundary may violate the original CRB requirement even when the corresponding power-allocation prediction error is small. To provide additional robustness, PATBO generates reference training labels using an adaptively tightened CRB requirement.

Let $C_{\mathrm{CRB}}^{\min}$ denote the minimum aggregate CRB
obtained under the physically constructed beam directions while
satisfying the communication-QoS, sensing-SINR, and transmit-power
requirements. Let $C_{\mathrm{CRB}}^{\mathrm{base}}$ denote the
aggregate CRB of the baseline structured beamformer. An adaptive
operational CRB threshold is first constructed as
\begin{equation}
\Gamma_{\mathrm{CRB}}^{\mathrm{orig}}
=
C_{\mathrm{CRB}}^{\min}
+
\eta_{\mathrm{CRB}}
\left(
C_{\mathrm{CRB}}^{\mathrm{base}}
-
C_{\mathrm{CRB}}^{\min}
\right),
\label{eq:original_crb_threshold}
\end{equation}
where the values $0.50$, $0.70$, $0.85$, $0.95$, and $1.00$ are
tested in order for $\eta_{\mathrm{CRB}}$ until a feasible reference solution is obtained.

To provide a safety margin for training while preventing the
tightened threshold from falling below the achievable CRB region,
the safe threshold is defined as
\begin{equation}
\Gamma_{\mathrm{CRB}}^{\mathrm{safe}}
=
\min\!\left\{
\Gamma_{\mathrm{CRB}}^{\mathrm{orig}},
\max\!\left[
\eta_s\Gamma_{\mathrm{CRB}}^{\mathrm{orig}},
(1+\epsilon_s)C_{\mathrm{CRB}}^{\min}
\right]
\right\},
\label{eq:safe_candidate}
\end{equation}
where $\eta_s=0.95$ is the CRB safety factor and $\epsilon_s=10^{-3}$ provides a small numerical buffer above the minimum achievable aggregate CRB $C_{\mathrm{CRB}}^{\min}$. The inner maximum prevents the tightened threshold from falling below the feasible CRB region, while the outer minimum ensures that the safe threshold does not exceed the original operational threshold. The complete safe-margin reference-generation procedure is summarized
in Algorithm~\ref{alg:safe_margin}.
\begin{algorithm}[t]
\caption{Safe-Margin Reference Generation}
\label{alg:safe_margin}
\begin{algorithmic}[1]
\REQUIRE $\mathbf{H}$, target positions, $P_t$, $\eta_s$,
$R_{\min}$, and $\gamma_{\min}^{s}$
\ENSURE $\boldsymbol{\rho}^{\star},
\Gamma_{\mathrm{CRB}}^{\mathrm{safe}}$

\STATE Construct the communication and sensing beam directions
according to \eqref{eq:rzf}--\eqref{eq:direction_matrix}.

\STATE Minimize the CRB over the stream powers subject to
communication QoS, sensing SINR, and transmit-power constraints
to obtain $C_{\mathrm{CRB}}^{\min}$.

\STATE Evaluate the aggregate CRB $C_{\mathrm{CRB}}^{\mathrm{base}}$ of the baseline structured beamformer.
\STATE Construct $\Gamma_{\mathrm{CRB}}^{\mathrm{orig}}$
according to \eqref{eq:original_crb_threshold}
\STATE Construct $\Gamma_{\mathrm{CRB}}^{\mathrm{safe}}$
according to \eqref{eq:safe_candidate}.
\STATE Solve the constrained reference power-allocation problem
under $\Gamma_{\mathrm{CRB}}^{\mathrm{safe}}$.
\STATE Normalize the optimized powers as
$\boldsymbol{\rho}^{\star}=\mathbf{p}^{\star}/P_t$.
\STATE \textbf{return}
$\boldsymbol{\rho}^{\star},
\Gamma_{\mathrm{CRB}}^{\mathrm{safe}}$.
\end{algorithmic}
\end{algorithm}

\subsection{CRB Sensitivity Estimation}

A small power-allocation error can have substantially different
physical consequences depending on the corresponding stream. PATBO
therefore estimates the local sensitivity of the aggregate CRB to
each normalized stream-power variable.

Let $C(\boldsymbol{\rho})$ denote the aggregate CRB obtained using
normalized power allocation $\boldsymbol{\rho}$. For stream $q$, a local simplex-constrained directional sensitivity is approximated using a central finite difference:
\begin{equation}
    g_{q,n}
    \approx
    \frac{
    C(\boldsymbol{\rho}_{q,n}^{+})
    -
    C(\boldsymbol{\rho}_{q,n}^{-})
    }{
    2\Delta\rho
    },
    \label{eq:crb_gradient}
\end{equation}
where $\Delta\rho=10^{-3}$.

Following each positive or negative perturbation, the power vector is projected back onto the unit simplex to preserve $\rho_q\geq0$ and $\sum_q\rho_q=1$. Consequently, $g_{q,n}$ should be interpreted as a coupled local directional sensitivity under the fixed-total-power simplex constraint, rather than as an unconstrained partial derivative with respect to a single independent power variable.
For numerical stability, the CRB gradient is normalized by the
corresponding safe CRB threshold:
\begin{equation}
    \widetilde{g}_{q,n}
    =
    \frac{
    g_{q,n}
    }{
    \Gamma_{\mathrm{CRB},n}^{\mathrm{safe}}
    }.
    \label{eq:normalized_gradient}
\end{equation}

\subsection{CRB-Sensitivity-Aware Training}
\label{subsec:sensitivity_training}

Let $\rho_{q,n}^{\star}$ denote the normalized reference stream power for entity $q$ in sample $n$, and let $\hat{\rho}_{q,n}$ denote the corresponding PATBO prediction. The stream-wise prediction error is
\begin{equation}
e_{q,n}
=
\hat{\rho}_{q,n}
-
\rho_{q,n}^{\star}.
\label{eq:prediction_error}
\end{equation}

Although a conventional mean-squared-error objective treats all power deviations equally, power-allocation errors can have substantially different physical consequences in near-field localization. In particular, a relatively small perturbation of a CRB-sensitive stream may cause a larger localization penalty than a comparable perturbation of a less-sensitive stream.

To reflect this behavior during training, the CRB-sensitivity weight is defined as
\begin{equation}
w_{q,n}^{\mathrm{sens}}
=
1
+
\lambda_s
\left|
\widetilde{g}_{q,n}
\right|,
\label{eq:sensitivity_weight}
\end{equation}
where $\widetilde{g}_{q,n}$ is the normalized simplex-constrained CRB sensitivity defined in \eqref{eq:normalized_gradient}, and
$\lambda_s=0.5$ controls the contribution of the physics-based weighting.

For a mini-batch containing $N_b$ samples, the proposed PATBO training objective is
\begin{equation}
\mathcal{L}_{\mathrm{PATBO}}
=
\frac{1}{N_b(K+L)}
\sum_{n=1}^{N_b}
\sum_{q=1}^{K+L}
w_{q,n}^{\mathrm{sens}}
\left(
\hat{\rho}_{q,n}
-
\rho_{q,n}^{\star}
\right)^2.
\label{eq:patbo_loss}
\end{equation}

Hence, the proposed objective assigns larger penalties to prediction errors associated with streams having stronger local impact on the aggregate localization CRB. The loss remains a weighted regression objective over the normalized stream powers, while the physically constructed communication and sensing beam directions are preserved outside the neural predictor.

The final PATBO formulation deliberately employs only the CRB-sensitivity weighting in the neural loss. The safe CRB threshold is retained during reference generation to construct robust training targets, but the safe-margin value is not used as an additional sample weight in the final training objective. The complete CRB-sensitivity estimation and PATBO training procedure
is summarized in Algorithm~\ref{alg:patbo_training}.
\begin{algorithm}[!t]
\caption{CRB-Sensitivity-Aware PATBO Training}
\label{alg:patbo_training}
\begin{algorithmic}[1]
\REQUIRE Training tokens $\{\mathbf{X}_n\}$,
reference powers $\{\boldsymbol{\rho}_n^{\star}\}$,
safe CRB thresholds
$\{\Gamma_{\mathrm{CRB},n}^{\mathrm{safe}}\}$,
perturbation $\Delta\rho$, and sensitivity factor $\lambda_s$
\ENSURE Trained PATBO parameters $\boldsymbol{\Theta}^{\star}$

\FOR{each training sample $n$}
    \FOR{$q=1,\ldots,K+L$}
        \STATE Perturb $\rho_{q,n}^{\star}$ by
        $\pm\Delta\rho$.
        \STATE Project the positive and negative perturbed power
        vectors onto the unit simplex.
        \STATE Evaluate the aggregate CRB for both perturbed
        allocations.
        \STATE Compute the finite-difference sensitivity using
        \eqref{eq:crb_gradient}.
        \STATE Normalize the sensitivity using
        \eqref{eq:normalized_gradient}.
        \STATE Compute $w_{q,n}^{\mathrm{sens}}$ using
        \eqref{eq:sensitivity_weight}.
    \ENDFOR
\ENDFOR

\REPEAT
    \STATE Predict normalized stream-power allocations for a
    mini-batch using the token-preserving Transformer.
    \STATE Compute $\mathcal{L}_{\mathrm{PATBO}}$ using
    \eqref{eq:patbo_loss}.
    \STATE Update the network parameters.
    \STATE Evaluate the validation loss.
\UNTIL{the validation stopping criterion is satisfied}

\RETURN Parameters corresponding to the best validation loss
\end{algorithmic}
\end{algorithm}

\subsection{Physical Feasibility Verification}
After reconstructing the Transformer-predicted beamformer
$\widehat{\mathbf{W}}$ according to \eqref{eq:beamformer_reconstruction}, its feasibility is evaluated using the original operational constraints rather than the tightened CRB threshold used for reference generation. For a candidate beamformer $\mathbf{W}$, define the feasibility indicator
\begin{equation}
\begin{split}
    \mathcal{F}(\mathbf{W})
    =
    \mathbb{I}
    \big\{&
    R_k\geq R_{\min},\ \forall k;
    \\
    &
    \gamma_l^s\geq\gamma_{\min}^s,\ \forall l;
    \\
    &
    C(\mathbf{W})
    \leq
    \Gamma_{\mathrm{CRB}}^{\mathrm{orig}};
    \\
    &
    \|\mathbf{W}\|_F^2\leq P_t
    \big\}.
\end{split}
\label{eq:feasibility_indicator}
\end{equation}
where $\mathcal{F}(\mathbf{W})=1$ indicates that $\mathbf{W}$
satisfies all original physical constraints, and $\mathcal{F}(\mathbf{W})=0$ otherwise. If $\mathcal{F}(\widehat{\mathbf{W}})=1$, the Transformer output is directly accepted; otherwise, physics-constrained repair is invoked.

\subsection{Physics-Constrained Power Repair}

If the Transformer prediction violates one or more original
constraints, PATBO invokes a low-dimensional physics-constrained
power-repair stage. The physical beam directions $\mathbf{V}$ remain fixed, while only the $K+L$ real-valued stream powers are optimized.

Let $\widehat{\mathbf{p}}$ denote the Transformer-predicted physical power vector. The repair problem is formulated as
\begin{subequations}
\label{eq:repair_problem}
\begin{align}
    \underset{\mathbf{p}}{\operatorname{minimize}}
    \quad&
    \left\|
    \frac{
    \mathbf{p}-\widehat{\mathbf{p}}
    }{
    P_t
    }
    \right\|_2^2
    \label{eq:repair_objective}
    \\
    \text{subject to}\quad&
    R_k(\mathbf{W}(\mathbf{p}))
    \geq
    R_{\min},
    \quad \forall k,
    \label{eq:repair_rate}
    \displaybreak[0]
    \\
    &
    \gamma_l^s(\mathbf{W}(\mathbf{p}))
    \geq
    \gamma_{\min}^{s},
    \quad \forall l,
    \label{eq:repair_sensing}
    \displaybreak[0]
    \\
    &
    C(\mathbf{W}(\mathbf{p}))
    \leq
    \Gamma_{\mathrm{CRB}}^{\mathrm{orig}},
    \label{eq:repair_crb}
    \\
    &
    \sum_{q=1}^{K+L}p_q
    =
    P_t,
    \label{eq:repair_power}
    \\
    &
    p_q\geq0,
    \quad
    q=1,\ldots,K+L.
    \label{eq:repair_nonnegative}
\end{align}
\end{subequations}
Here, $\mathbf{W}(\mathbf{p})=\mathbf{V}\operatorname{diag}(\sqrt{\mathbf{p}})$.

The repair problem therefore operates over only $K+L$ real-valued
stream-power variables, while the physically constructed beam
directions remain fixed. Consequently, the dimensionality of the
repair optimization is independent of the number of antenna elements $M$.

The first repair attempt is initialized as
$\mathbf{p}^{(0)}=\widehat{\mathbf{p}}$. If this attempt fails to
recover feasibility, a second sensing-oriented initialization is
employed. Let $P_c=\sum_{k=1}^{K}\hat{p}_k$ denote the communication-stream power. A fraction $\delta_s=0.1$ of $P_c$ is removed proportionally from the communication streams and distributed equally among the $L$ sensing streams. The repair problem is then solved again from this modified initialization.

In the implementation, the constrained power-repair problem is solved using sequential quadratic programming (SQP), a standard numerical approach for constrained nonlinear optimization
\cite{Nocedal2006Numerical}. Since repair is invoked only for
infeasible Transformer predictions, its computational cost is avoided whenever direct neural inference already produces a feasible physical solution.

\subsection{Hybrid PATBO Inference}

Although physics-constrained repair can recover a substantial
portion of infeasible Transformer predictions, a repair solution is
not assumed to exist for every realization. Therefore, the final PATBO framework includes a constrained optimizer fallback for cases that remain infeasible after physics-based repair.

Let $\mathbf{W}_{\mathrm{rep}}$ denote the repaired solution and
$\mathbf{W}_{\mathrm{fb}}$ denote a feasible solution returned by the constrained optimizer fallback. The final hybrid beamformer is selected according to
\begin{equation}
\mathbf{W}_{\mathrm{PATBO}}
=
\begin{cases}
\widehat{\mathbf{W}},
&
\mathcal{F}(\widehat{\mathbf{W}})=1,
\\[2mm]
\mathbf{W}^{\mathrm{rep}},
&
\mathcal{F}(\widehat{\mathbf{W}})=0
\ \text{and}\
\mathcal{F}(\mathbf{W}^{\mathrm{rep}})=1,
\\[2mm]
\mathbf{W}^{\mathrm{fb}},
&
\text{otherwise}.
\end{cases}
\label{eq:hybrid_patbo}
\end{equation}

In the offline retained-test-set evaluation, the stored feasible
reference solution is used as the fallback realization. Therefore, the reported $100\%$ final test-set feasibility should be interpreted as an empirical result on the retained evaluation set rather than as a universal guarantee that every unseen realization can be recovered by the proposed direct or repair stages.

Hence, PATBO follows a hierarchical computational strategy.
Transformer inference provides the lowest-latency solution.
Physics-constrained power repair is activated only when necessary,
while the reference optimizer is reserved for scenarios that cannot be recovered through low-dimensional repair. The resulting architecture therefore provides multiple operating modes with different latency--reliability tradeoffs. Direct Transformer inference offers the lowest online latency, whereas physics-constrained repair and reference optimization can be invoked when stronger feasibility guarantees are required.

An additional advantage of the structured formulation is its favorable scaling with the array size. Although increasing $M$ enlarges the physical communication and sensing beam directions, the Transformer continues to predict only $K+L$ stream-power variables. Hence, for fixed numbers of users and targets, the learned output dimension is independent of $M$. This property is particularly relevant to E-MIMO systems, where direct prediction of the complete beamforming matrix becomes increasingly challenging as the array size grows. The proposed structure therefore retains the array-dependent near-field propagation characteristics in the physical beam construction while confining learning to the lower-dimensional resource-allocation problem.

\section{Simulation Setup and Numerical Results}
\label{sec:results}

\subsection{Simulation Setup}

The proposed PATBO framework is evaluated using a $32\times32$ UPA with $M=1024$ antennas, $K=4$ communication users, and $L=2$ sensing targets at $f_c=8$~GHz. The antenna spacing is $\lambda/2$, and the users and targets are randomly located in the radiative near-field region at ranges of $8$--$12$~m. The target-position uncertainty along each Cartesian dimension is independently generated between $0.01$ and $0.10$~m. The communication and sensing beam directions are constructed using RZF and spherical-wave near-field focusing, respectively. The remaining system parameters are summarized in Table~\ref{tab:simulation_parameters}.

\begin{table}[!t]
\centering
\caption{Simulation parameters}
\label{tab:simulation_parameters}
\begin{tabular}{lc}
\hline
\textbf{Parameter} & \textbf{Value} \\
\hline
Number of BS antennas, $M$ & 1024 \\
UPA configuration & $32\times32$ \\
Number of communication users, $K$ & 4 \\
Number of sensing targets, $L$ & 2 \\
Carrier frequency, $f_c$ & 8 GHz \\
Antenna spacing, $d$ & $\lambda/2$ \\
Total normalized transmit power, $P_t$ & 100 \\
Noise power, $\sigma_n^2$ & 1 \\
Minimum user rate, $R_{\min}$ & 1 bit/s/Hz \\
Minimum sensing SINR & 5 dB \\
User/target range & 8--12 m \\
Target-position uncertainty & 0.01--0.10 m \\
CRB safety factor, $\eta_s$ & 0.95 \\
Generated scenarios & 5000 \\
Retained valid scenarios & 4632 \\
Training/validation/test split & 3242/694/696 \\
\hline
\end{tabular}
\end{table}

A total of 5000 random scenarios are generated. A realization is
retained only when the structured reference optimization produces
a valid solution satisfying the imposed physical constraints.
This results in 4632 valid realizations ($92.64\%$), which are
divided into 3242 training, 694 validation, and 696 test samples.
The safe-margin procedure of Section~III-F uses $\eta_s=0.95$; over the retained dataset, the mean reference, safe-threshold, and original-threshold CRBs are $185.3078$, $186.4777$, and $191.8442$, respectively.

The geometry-aware tokens are constructed according to \eqref{eq:comm_token}--\eqref{eq:transformer_input}, with feature
standardization based exclusively on the training set. The final
token-preserving PATBO configuration and training parameters are
summarized in Table~\ref{tab:patbo_parameters}.

\begin{table}[t]
\centering
\caption{PATBO network and training parameters}
\label{tab:patbo_parameters}
\begin{tabular}{lc}
\hline
\textbf{Parameter} & \textbf{Value} \\
\hline
Input features per token, $D_{\mathrm{in}}$ & 10 \\
Number of tokens, $K+L$ & 6 \\
Model dimension, $d_{\mathrm{model}}$ & 64 \\
Attention heads & 4 \\
Feed-forward dimension & 128 \\
Transformer encoder layers & 2 \\
Token-wise hidden dimension & 64 \\
Dropout probability & 0.1 \\
Mini-batch size & 32 \\
Maximum epochs & 100 \\
Initial learning rate & $10^{-3}$ \\
Early-stopping patience & 15 \\
CRB sensitivity factor, $\lambda_s$ & 0.5 \\
Finite-difference perturbation, $\Delta\rho$ & $10^{-3}$ \\
\hline
\end{tabular}
\end{table}

\subsection{Training and Power-Prediction Performance}
\label{subsec:training_performance}

The final token-preserving PATBO model is trained using the
CRB-sensitivity-aware objective in \eqref{eq:patbo_loss}. Early stopping terminates training after $51$ epochs, with the best validation loss of $3.7120\times10^{-3}$ obtained at epoch $36$.

On the independent $696$-sample test set, the predicted normalized
stream-power allocations achieve a mean-squared error of $3.1727\times10^{-3}$. Using the common sample-wise normalized-error
metric adopted for the learning-baseline comparison, the corresponding power-allocation NMSE is $-11.1670$~dB. The softmax output satisfies the unit-simplex constraint numerically, with a maximum power-sum error of
$3.33\times10^{-16}$.

\begin{figure}[t]
\centering
\includegraphics[width=\columnwidth]{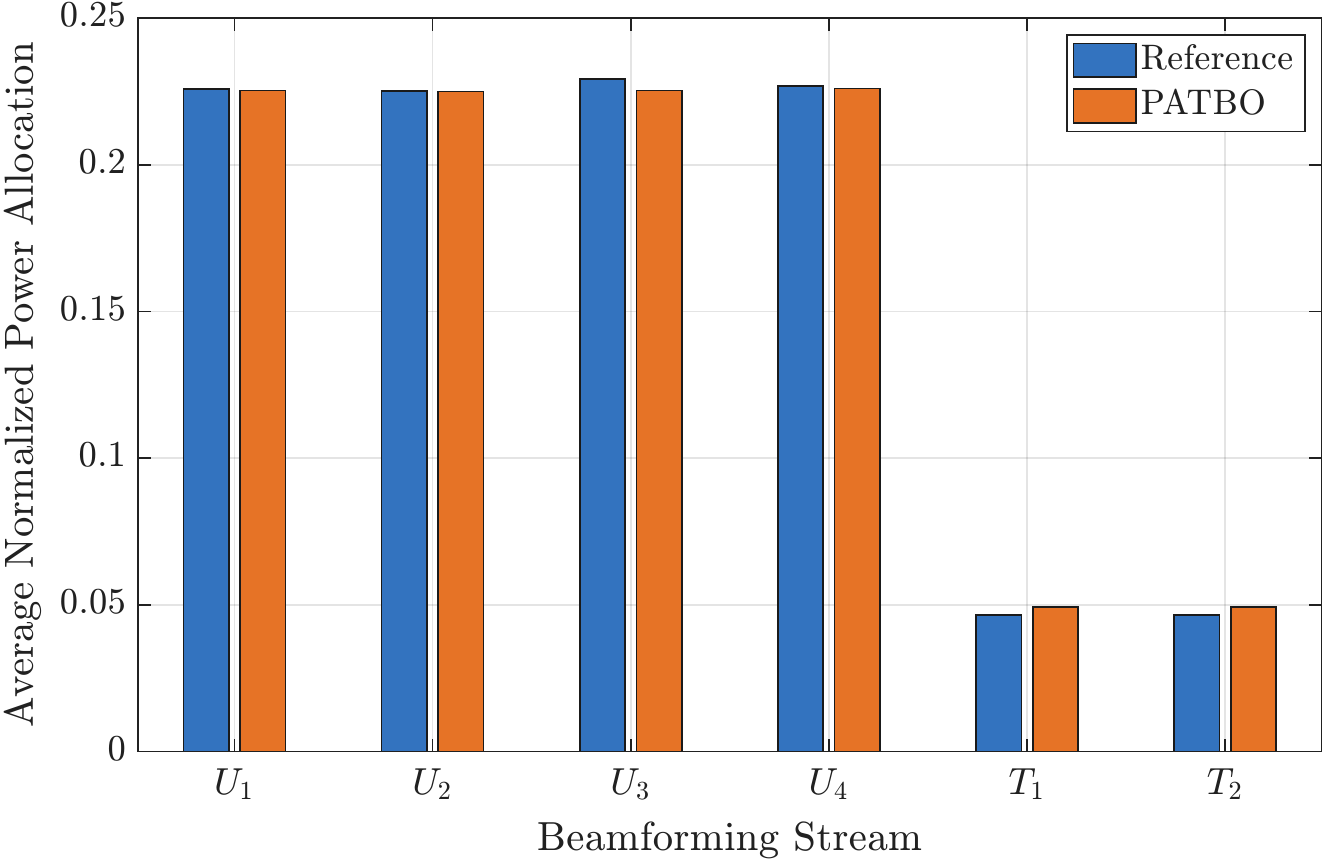}
\caption{Average normalized stream-power allocation of the reference
solution and PATBO over the 696 retained test samples.}
\label{fig:average_power}
\end{figure}

Fig.~\ref{fig:average_power} compares the average normalized
stream-power allocation of PATBO with the reference solution.
PATBO closely reproduces the overall communication--sensing power
distribution, with most of the available power assigned to the
communication streams and a smaller fraction allocated to sensing.

Nevertheless, agreement in average power allocation or conventional NMSE does not by itself imply physical feasibility. The localization CRB is highly sensitive to certain coupled perturbations of the stream-power vector, and therefore relatively small prediction errors can move a realization outside the feasible CRB region. This motivates the explicit constraint evaluation presented next.

\subsection{Communication and Sensing Performance}

Table~\ref{tab:direct_patbo_performance} summarizes the performance of direct token-preserving PATBO before feasibility recovery. The reference and PATBO achieve average communication sum rates of $17.9204$ and $18.0966$~bit/s/Hz, respectively, corresponding to $100.98\%$ rate retention. The slightly higher learned rate should not be interpreted as outperforming the constrained reference, since some PATBO predictions violate the localization constraint.

The reference and PATBO aggregate CRBs are $181.9202$ and
$199.9954$, respectively, corresponding to a $9.94\%$ increase.
Communication QoS and transmit-power feasibility are $100\%$, and
sensing-SINR feasibility is $99.86\%$. In contrast, only $19.54\%$
of the direct predictions satisfy the original CRB constraint, which also limits the overall joint feasibility to $19.54\%$. The tighter safe CRB threshold is satisfied by $13.79\%$ of the predictions.

These results identify localization CRB as the dominant feasibility
bottleneck. Despite an NMSE of $-11.1670$~dB and near-reference
communication performance, small errors in the allocated powers can
significantly affect the FIM and CRB, both of which depend on geometry. Explicit post-inference feasibility verification is therefore required.

\begin{table}[t]
\centering
\caption{Performance and feasibility of direct PATBO}
\label{tab:direct_patbo_performance}
\begin{tabular}{lcc}
\hline
\textbf{Metric} & \textbf{Reference} & \textbf{PATBO} \\
\hline
Sum rate (bit/s/Hz)       & 17.9204 & 18.0966 \\
Rate retention (\%)       & 100.00  & 100.98 \\
Aggregate CRB             & 181.9202 & 199.9954 \\
CRB change (\%)           & --      & +9.94 \\
Communication QoS (\%)    & 100.00  & 100.00 \\
Sensing SINR (\%)         & 100.00  & 99.86 \\
Original CRB feasibility (\%) & 100.00 & 19.54 \\
Safe CRB feasibility (\%) & 100.00 & 13.79 \\
Power feasibility (\%)    & 100.00 & 100.00 \\
Joint feasibility (\%)    & 100.00 & 19.54 \\
\hline
\end{tabular}
\end{table}

\subsection{Constraint Feasibility and Physics-Constrained Repair}

Physics-constrained repair is applied to the 560 direct PATBO
predictions that violate at least one original constraint. The repair stage preserves the RZF communication and near-field sensing
directions and optimizes only the $K+L=6$ stream-power variables
according to \eqref{eq:repair_problem}.

Of the 560 repair attempts, 475 successfully recover feasibility,
corresponding to a conditional success rate of $84.82\%$. Together
with the 136 directly feasible PATBO predictions, this increases
joint feasibility from $19.54\%$ to $87.79\%$, as summarized in
Table~\ref{tab:repair_feasibility}. The average relative power
adjustment among successfully repaired samples is $17.403\%$. Thus, repair recovers most direct failures without altering the physics-based beam directions.

Repair also improves the localization performance. The average
aggregate CRB decreases from $199.9954$ for direct PATBO to
$190.5222$ after direct acceptance and repair, reducing the CRB
degradation relative to the reference from $9.94\%$ to $4.73\%$.
The corresponding average sum rate is $17.9573$~bit/s/Hz
($100.21\%$ reference-rate retention), while the power-allocation
NMSE improves from $-11.1670$ to $-13.5352$~dB.

Nevertheless, 85 of the 560 repair attempts remain infeasible.
These realizations, corresponding to $12.21\%$ of the complete
test set, are subsequently handled by the optimizer-fallback stage.

\begin{table}[t]
\centering
\caption{Constraint feasibility before and after repair.}
\label{tab:repair_feasibility}
\begin{tabular}{lcc}
\hline
\textbf{Constraint} & \textbf{Direct PATBO} & \textbf{After Repair} \\
\hline
Communication QoS (\%) & 100.00 & 100.00 \\
Sensing SINR (\%)      & 99.86  & 100.00 \\
CRB (\%)               & 19.54  & 87.79 \\
Power (\%)             & 100.00 & 100.00 \\
Joint feasibility (\%) & 19.54  & 87.79 \\
\hline
\end{tabular}
\end{table}

\subsection{Hybrid PATBO Performance}
\label{subsec:hybrid_performance}

\begin{figure}[t]
\centering
\includegraphics[width=\columnwidth]
{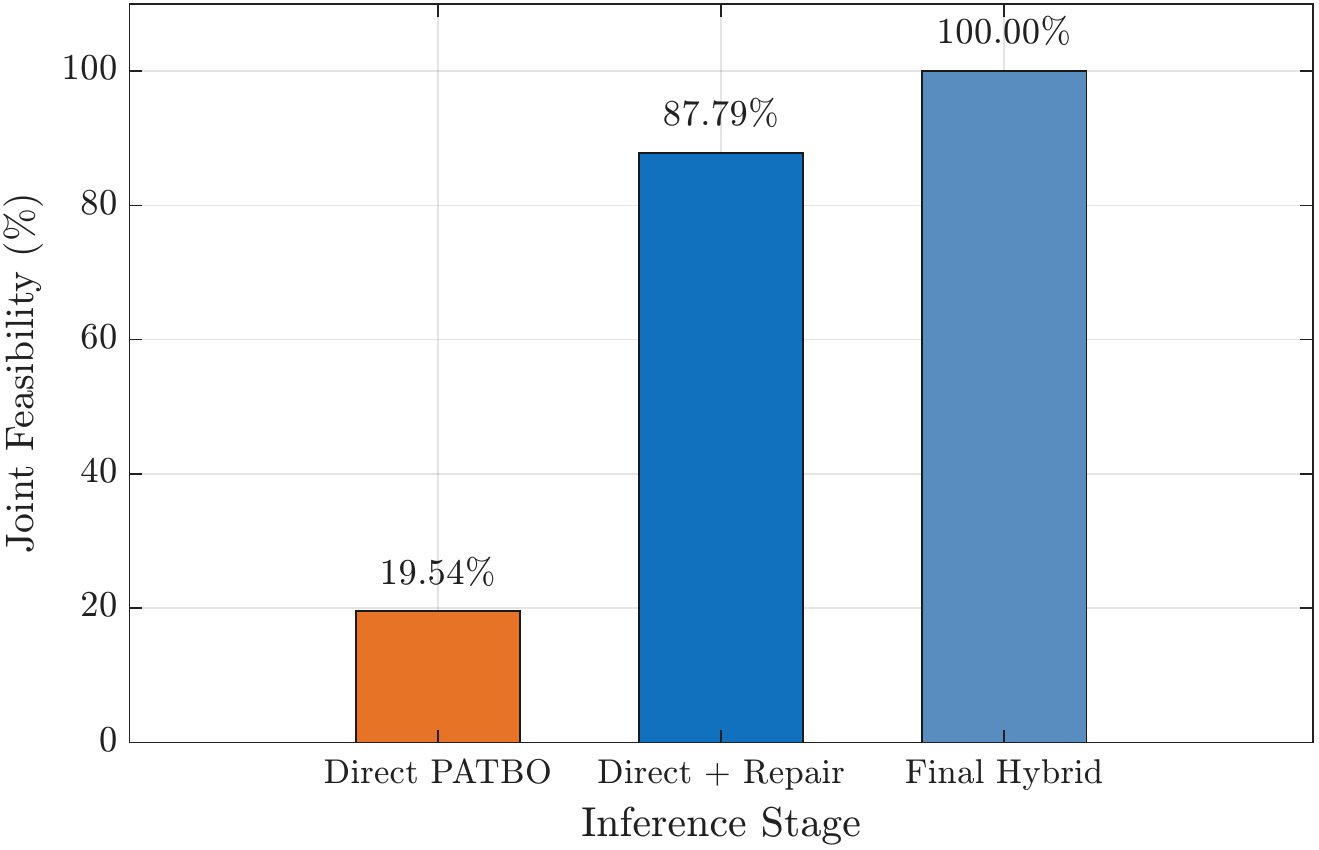}
\caption{Joint-feasibility progression across direct PATBO, physics-constrained repair, and optimizer fallback.}
\label{fig:hybrid_feasibility}
\end{figure}

The complete hybrid PATBO framework combines direct inference,
physics-constrained repair, and optimizer fallback. Among the 696
retained test realizations, 136 are directly feasible, 475 additional realizations are recovered through repair, and the remaining 85 require fallback. Consequently, joint feasibility progresses from $19.54\%$ for direct PATBO to $87.79\%$ after repair and finally to $100\%$ after fallback, as illustrated in
Fig.~\ref{fig:hybrid_feasibility}.

The final hybrid beamformer achieves an average sum rate of
$17.9004$~bit/s/Hz, corresponding to $99.89\%$ reference-rate
retention. Its average aggregate CRB is $184.4167$, only $1.37\%$
above the reference value of $181.9202$. Thus, the CRB degradation
decreases from $9.94\%$ for direct PATBO to $4.73\%$ after repair
and to $1.37\%$ for the complete hybrid output.

\begin{figure}[!t]
\centering
\includegraphics[width=\columnwidth]
{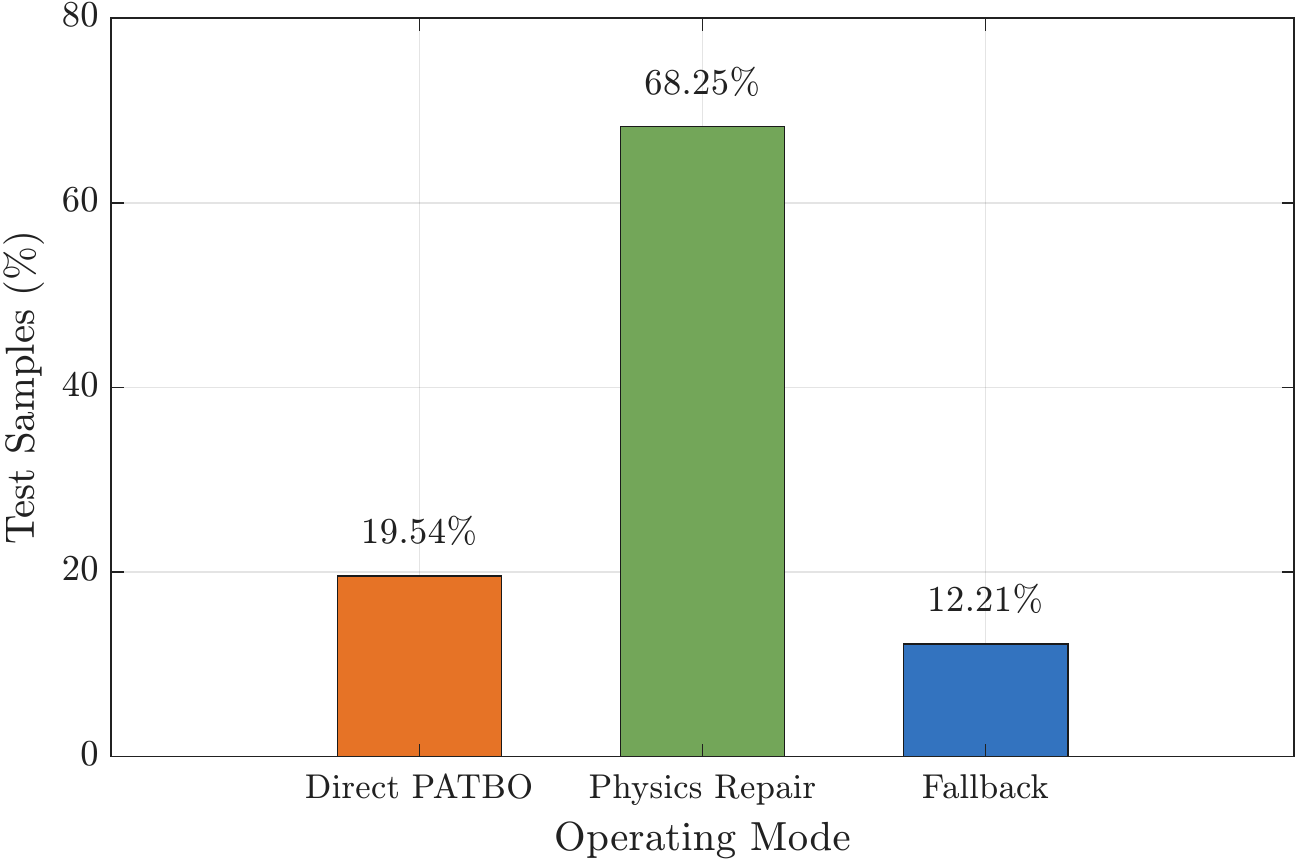}
\caption{Distribution of PATBO inference operating modes over the
696-sample test set.}
\label{fig:patbo_operating_modes}
\end{figure}

The corresponding operating-mode distribution is shown in
Fig.~\ref{fig:patbo_operating_modes}. Direct PATBO provides the final solution for $19.54\%$ of the samples, successful repair handles an additional $68.25\%$, and only $12.21\%$ require optimizer fallback. Hence, $87.79\%$ of the retained test realizations are handled without invoking the fallback optimizer.

The reported $100\%$ final feasibility is an empirical result on
the retained test set rather than a universal guarantee. For the
85 realizations that remain infeasible after repair, the offline
evaluation uses the stored feasible reference solution; an online
implementation would instead invoke the constrained optimizer.

\subsection{Computational Complexity and Runtime Analysis}
\label{subsec:runtime}

PATBO replaces repeated numerical optimization with direct prediction
of the $K+L=6$ stream-power variables while retaining the physically
constructed beam directions. As established in Section III, this
learned output dimension is independent of the antenna-domain
beamformer dimension for fixed $K$ and $L$.

\begin{figure}[!t]
\centering
\includegraphics[width=\columnwidth]
{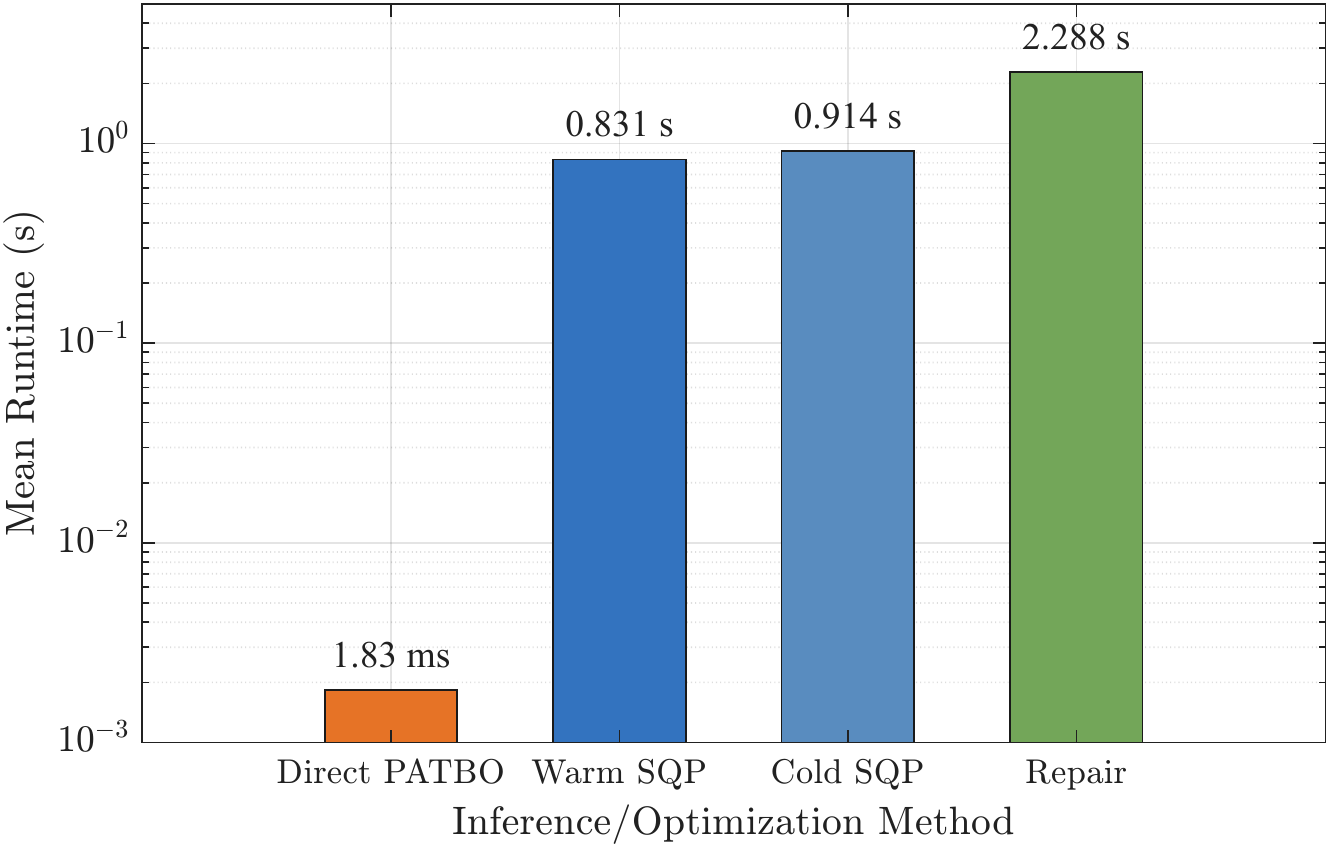}
\caption{Mean per-sample runtime of PATBO inference, SQP, and physics-constrained repair.}
\label{fig:runtime_comparison}
\end{figure}

Runtime is evaluated using the actual PATBO inference, SQP, and repair
routines. Single-sample PATBO latency is measured after warm-up and
averaged over repeated inference trials, while batched throughput is
measured over the complete test set. Cold-start and PATBO-warm-start
SQP are evaluated on a deterministic 100-sample subset using equal-power
and PATBO-predicted initializations, respectively. The measured results
for $M=1024$ are summarized in Table \ref{tab:runtime_comparison} and Fig. \ref{fig:runtime_comparison}.

Direct PATBO requires an average single-sample inference latency of
$1.1085$~ms, compared with $0.9512$~s for cold-start SQP and
$0.8641$~s for PATBO-warm-start SQP. Thus, direct PATBO is approximately
$858\times$ and $779\times$ faster than cold-start and PATBO-warm-start
SQP, respectively. Batched PATBO further reduces the average processing
time to $0.0367$~ms/sample, although this value represents batch
throughput rather than single-sample online latency. Physics-constrained
repair requires $2.4011$~s on average and is therefore employed
selectively for feasibility recovery.

The outcome column in Table~V complements the runtime comparison.
Cold-start and PATBO-warm-start SQP achieve $39\%$ and $50\%$
feasibility, respectively, indicating that PATBO predictions can provide
a useful initialization for numerical optimization. Physics-constrained
repair successfully recovers $84\%$ of the directly infeasible samples
in the runtime benchmark. These SQP results represent the evaluated
local numerical solver rather than a globally optimal or universally
feasible optimizer. Overall, the results support the proposed hierarchy:
low-latency direct PATBO inference is attempted first, while iterative
physics-based optimization is invoked selectively when additional
feasibility recovery is required.

\begin{table}[!t]
\centering
\caption{Measured Runtime of PATBO and Numerical Optimization}
\label{tab:runtime_comparison}
\begin{tabular}{lcc}
\hline
\textbf{Method} & \textbf{Mean Runtime} & \textbf{Outcome} \\
\hline
PATBO, single sample & $1.1085$ ms & Direct prediction \\
PATBO, batched       & $0.0367$ ms/sample & Throughput \\
Cold-start SQP       & $0.9512$ s & $39\%$ feasible \\
PATBO-warm SQP       & $0.8641$ s & $50\%$ feasible \\
Physics repair       & $2.4011$ s & $84\%$ success \\
\hline
\end{tabular}
\end{table}

\begin{table*}[!t]
\centering
\caption{Direct-inference ablation and baseline comparison}
\label{tab:ablation_comparison}
\begin{tabular}{lccccc}
\hline
\textbf{Method}
& \textbf{NMSE (dB)}
& \textbf{Rate Ret. (\%)}
& \textbf{CRB Change (\%)}
& \textbf{Sensing (\%)}
& \textbf{Joint Feas. (\%)} \\
\hline
MLP
& $-10.7953$ & $101.18$ & $+11.47$ & $98.13$ & $13.94$ \\
Vanilla Transformer
& $-10.8379$ & $101.14$ & $+12.05$ & $99.71$ & $10.06$ \\
Pooled sensitivity-aware Transformer
& $-10.8144$ & $101.28$ & $+11.98$ & $99.57$ & $11.64$ \\
Pooled margin-only Transformer
& $-10.8379$ & $101.07$ & $+11.94$ & $99.71$ & $9.20$ \\
Pooled sensitivity+margin Transformer
& $-10.8368$ & $100.93$ & $+11.83$ & $99.71$ & $8.48$ \\
\textbf{Token-preserving sensitivity-aware PATBO}
& $\mathbf{-11.1670}$ & $100.98$ & $\mathbf{+9.94}$
& $\mathbf{99.86}$ & $\mathbf{19.54}$ \\
\hline
\end{tabular}
\end{table*}

\subsection{Scalability with E-MIMO Array Size}

To evaluate scalability, we vary the number of BS antennas while
keeping $K=4$ communication users and $L=2$ sensing targets fixed.
Four square UPA configurations, $20\times20$, $24\times24$,
$28\times28$, and $32\times32$, are considered, corresponding to
$M=400$, $576$, $784$, and $1024$, respectively. For each configuration,
a new dataset is generated and PATBO is retrained using the same network
architecture and training procedure. The corresponding Rayleigh
distances are $13.54$, $19.84$, $27.34$, and $36.04$~m, maintaining
the considered $8$--$12$~m operating region within the near-field regime.

\begin{figure}[!t]
\centering
\includegraphics[width=\columnwidth]
{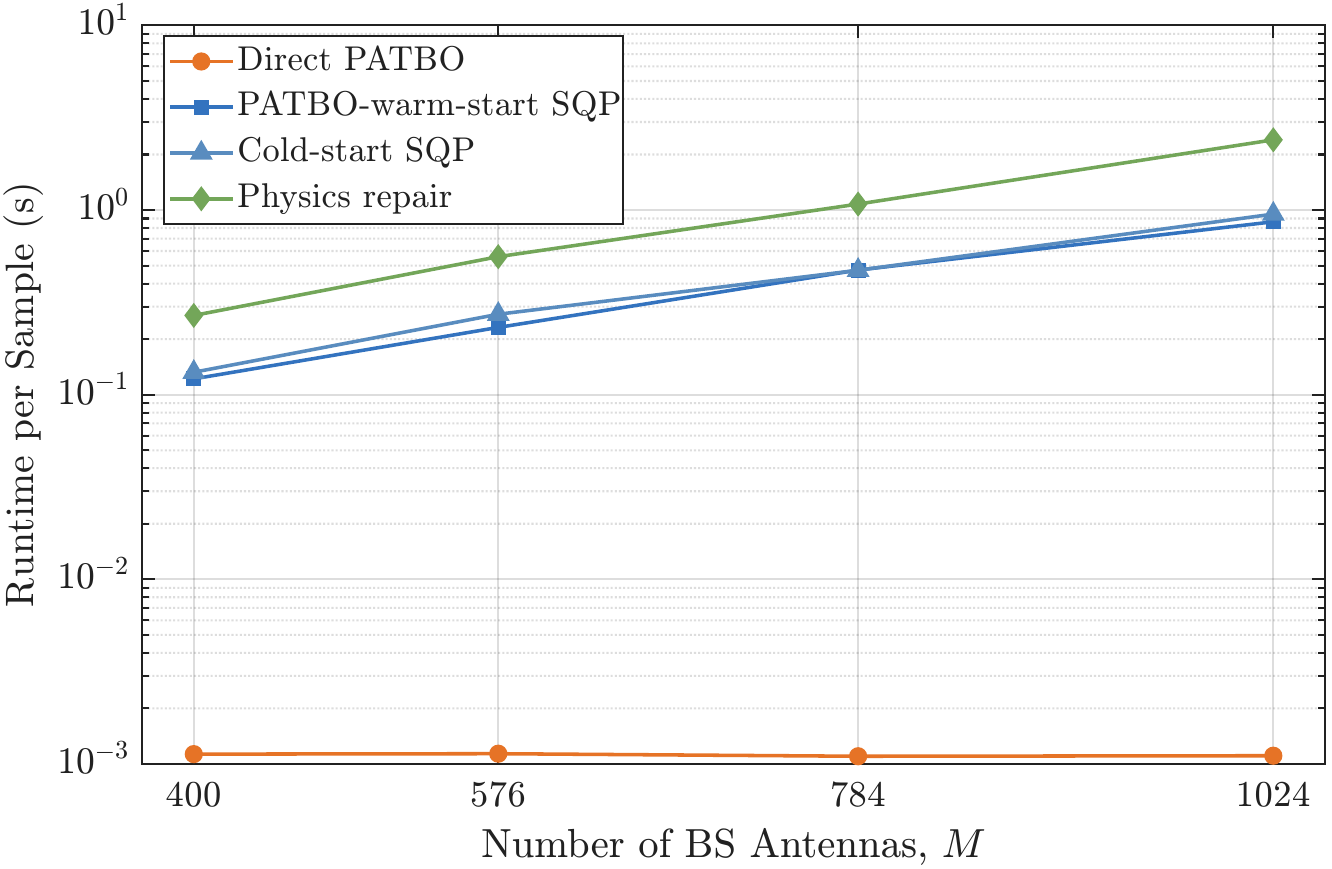}
\caption{Runtime scalability of the proposed PATBO framework versus the
number of BS antenna elements $M$, including direct PATBO inference,
PATBO-warm-start SQP, cold-start SQP, and physics-constrained repair.}
\label{fig:scalability_runtime}
\end{figure}

Fig.~\ref{fig:scalability_runtime} shows the measured runtime as a
function of $M$. Direct PATBO exhibits nearly constant inference
latency, with mean runtimes of $1.131$, $1.137$, $1.101$, and
$1.109$~ms for $M=400$, $576$, $784$, and $1024$, respectively.
This behavior is enabled by the structured representation, where the
Transformer predicts only $K+L=6$ stream-power variables irrespective
of the number of antenna elements. In contrast, cold-start SQP runtime
increases from $0.132$~s to $0.951$~s and PATBO-warm-start SQP from
$0.122$~s to $0.864$~s as $M$ increases from $400$ to $1024$.
Similarly, the physics-constrained repair runtime increases from
$0.269$~s to $2.401$~s.

The reference-rate retention remains above $100\%$ for all evaluated
array sizes. Although direct joint feasibility varies with the array
configuration, physics-constrained repair substantially improves
feasibility, while the complete verify--repair--fallback framework
achieves $100\%$ final feasibility for all considered values of $M$.
Since repair is invoked only for samples failing the direct feasibility
check, its increasing runtime does not affect direct PATBO inference
latency. Overall, the results demonstrate favorable computational
scalability of the proposed framework for large-aperture near-field
E-MIMO-ISAC systems.

\subsection{Ablation and Baseline Analysis}
\label{subsec:ablation}

To isolate the contributions of the proposed architecture and training objective, token-preserving sensitivity-aware PATBO is compared with an MLP and four Transformer variants under the same dataset, test split, and structured beamformer reconstruction. Table~\ref{tab:ablation_comparison} summarizes their direct-inference performance before repair or optimizer fallback.

\begin{figure}[!t]
    \centering
    \includegraphics[width=\columnwidth]
    {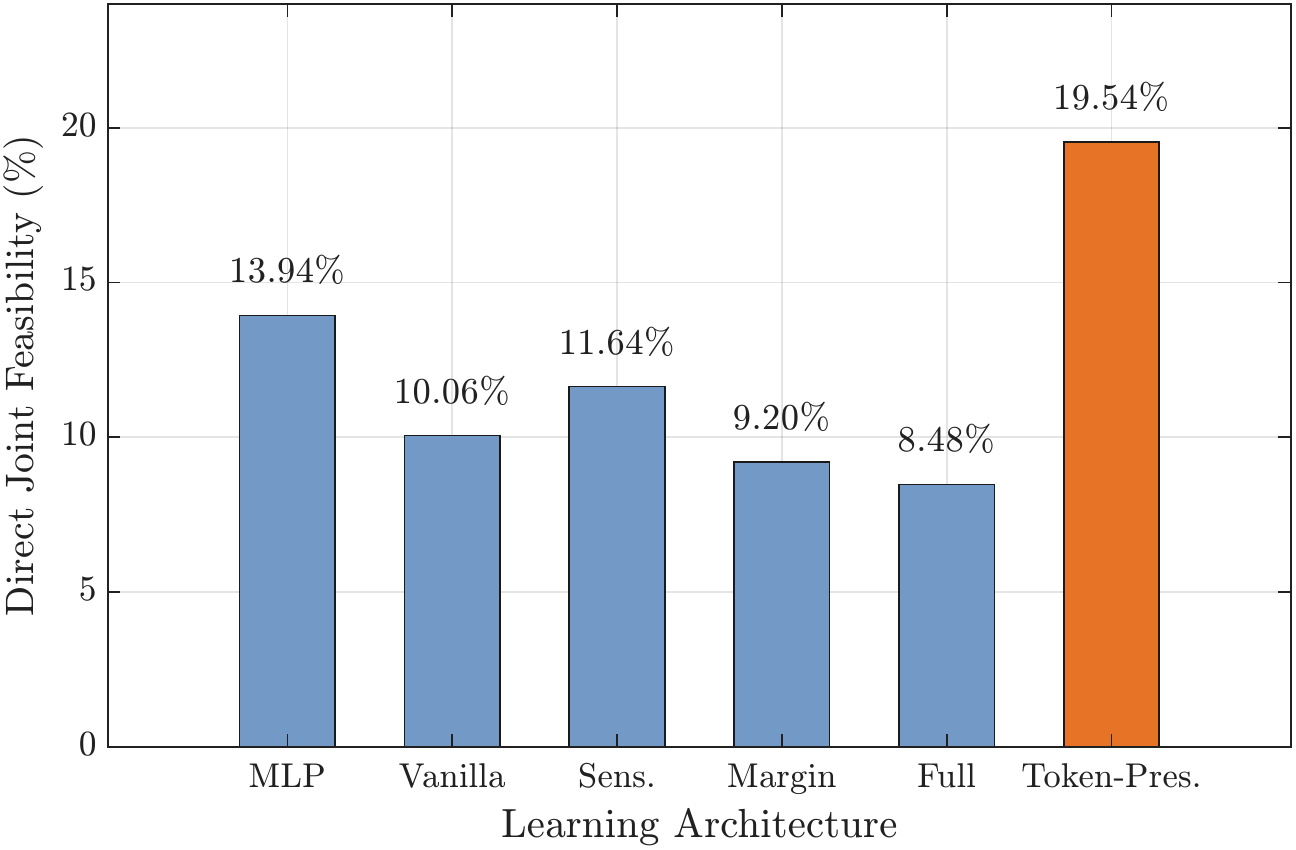}
    \caption{Direct joint-feasibility comparison among the MLP baseline,
    Transformer ablations, and the proposed token-preserving
    sensitivity-aware PATBO over the 696-sample test set.}
    \label{fig:ablation_feasibility}
\end{figure}

The proposed token-preserving PATBO achieves the highest direct joint feasibility of $19.54\%$, compared with $13.94\%$ for the MLP and $11.64\%$ for the pooled sensitivity-aware Transformer, as highlighted in Fig.~\ref{fig:ablation_feasibility}. Relative to the latter, this represents a $7.90$ percentage-point improvement. PATBO also achieves the lowest power-allocation NMSE of $-11.1670$~dB and reduces the average CRB degradation from $11.98\%$ for the pooled sensitivity-aware model to $9.94\%$.

The improvement is consistent with preserving the association
between each contextualized entity token and its corresponding power variable. In contrast, the pooled Transformer variants compress the entity tokens into a single representation before generating the stream powers, thereby removing this explicit correspondence.

The loss-function ablation further shows that sensitivity-aware
training is more effective than margin-based weighting for the
considered dataset. Within the pooled architecture, sensitivity-aware, margin-only, and combined sensitivity-and-margin training achieve $11.64\%$, $9.20\%$, and $8.48\%$ joint feasibility, respectively. Accordingly, the final PATBO model uses the CRB-sensitivity-aware objective with $\lambda_s=0.5$, while the safe CRB threshold is confined to reference-solution generation.

Although all learning models retain approximately $100\%$ of the
reference communication rate, their CRB feasibility differs
substantially. This confirms that communication-rate retention alone is insufficient for architecture selection and motivates the
verify--repair--fallback mechanism used by PATBO.

\let\PATBOFloatBarrier\FloatBarrier
\let\FloatBarrier\relax
\section{Conclusion}
\let\FloatBarrier\PATBOFloatBarrier
\label{sec:conclusion}

This paper proposed PATBO, a physics-aware Transformer framework for
near-field E-MIMO-ISAC beamforming that combines physically structured
beam directions with learned stream-power allocation. The
token-preserving architecture and CRB-sensitivity-aware training enable
low-dimensional power prediction while accounting for
localization-sensitive constraints. Numerical results show that
physics-constrained repair increases joint feasibility from 19.54\% for
direct PATBO to 87.79\%, while the complete verify--repair--fallback
framework achieves 99.89\% reference-rate retention and 100\% feasibility
on the retained test set. Direct PATBO inference requires only
1.11~ms per sample, corresponding to approximately 858$\times$ and
779$\times$ lower runtime than the evaluated cold-start and
PATBO-warm-start SQP benchmarks, respectively. Moreover, the scalability
results show that PATBO maintains approximately 1.1~ms inference latency
across $M=400$--$1024$, while the numerical optimization stages exhibit
increasing runtime with array size. These results demonstrate the
potential of combining structured learning with selective physics-based
optimization for low-latency and scalable near-field E-MIMO-ISAC
beamforming. Future work will investigate dynamic mobility, channel
uncertainty, and real-time implementation.

\bibliographystyle{IEEEtran}
\bibliography{PATBO_references}
\end{document}